\documentclass[aps,prd,floatfix,superscriptaddress,preprintnumbers,nofootinbib,preprint,tightenlines,longbibliography]{revtex4-1}
\pdfoutput=1
\usepackage[english]{babel}
\usepackage{amsmath,amssymb,amsbsy,amstext,amsthm,booktabs,exscale,relsize,slashed,graphicx,amsfonts,upgreek,xcolor}
\usepackage{multirow,dcolumn,bm,enumerate,url,hyperref}
\definecolor{lightsaberblue}{rgb}{.0,.1,.4}

\hypersetup{
  colorlinks   = true, 
  urlcolor     = lightsaberblue, 
  linkcolor    = lightsaberblue, 
  citecolor   = lightsaberblue 
}

\newcommand{\kev}{{\rm keV}}
\newcommand{\gev}{{\rm GeV}}
\newcommand{\mev}{{\rm MeV}}

\newcommand{\acro}[1]{\textsc{\MakeLowercase{#1}}}

\begin{document}
\title{\bf Accelerating Composite Dark Matter Discovery with Nuclear Recoils and the Migdal Effect}

\author{Javier F.  Acevedo}
\thanks{{\scriptsize Email}: \href{mailto:17jfa1@queensu.ca}{17jfa1@queensu.ca}; {\scriptsize ORCID}: \href{http://orcid.org/0000-0003-3666-0951}{0000-0003-3666-0951}}
\affiliation{\smaller The Arthur B. McDonald Canadian Astroparticle Physics Research Institute, \protect\\ Department of Physics, Engineering Physics, and Astronomy, \protect\\ Queen's University, Kingston, Ontario, K7L 2S8, Canada}

\author{Joseph Bramante}
\thanks{{\scriptsize Email}: \href{mailto:joseph.bramante@queensu.ca}{joseph.bramante@queensu.ca}; {\scriptsize ORCID}: \href{http://orcid.org/0000-0001-8905-1960}{0000-0001-8905-1960}}
\affiliation{\smaller The Arthur B. McDonald Canadian Astroparticle Physics Research Institute, \protect\\ Department of Physics, Engineering Physics, and Astronomy, \protect\\ Queen's University, Kingston, Ontario, K7L 2S8, Canada}
\affiliation{\smaller Perimeter Institute for Theoretical Physics, Waterloo, ON N2J 2W9, Canada}

\author{Alan Goodman}
\thanks{{\scriptsize Email}: \href{mailto:alan.goodman@queensu.ca}{alan.goodman@queensu.ca}; {\scriptsize ORCID}: \href{http://orcid.org/0000-0001-5289-258X}{0000-0001-5289-258X}}
\affiliation{\smaller The Arthur B. McDonald Canadian Astroparticle Physics Research Institute, \protect\\ Department of Physics, Engineering Physics, and Astronomy, \protect\\ Queen's University, Kingston, Ontario, K7L 2S8, Canada}

\begin{abstract}
Large composite dark matter states source a scalar binding field that, when coupled to Standard Model nucleons, provides a potential under which nuclei recoil and accelerate to energies capable of ionization, radiation, and thermonuclear reactions. We show that these dynamics are detectable for nucleon couplings as small as $g_n \sim 10^{-17}$ at dark matter experiments, where the greatest sensitivity is attained by considering the Migdal effect. We also explore Type-Ia supernovae and planetary heating as possible means to discover this type of dark matter.
\end{abstract}
\maketitle

\newpage
\tableofcontents

\section{Introduction}
\label{sec:int}

The existence of dark matter is well established through its gravitational interactions with visible matter. However, dark matter's cosmological formation, mass, and non-gravitational interactions remain unknown. An intriguing possibility is that dark matter forms bound states in the early universe, much like nuclei form during big bang nucleosynthesis, in a scenario often dubbed ``composite dark matter."

A simple model of composite dark matter that has been studied in \cite{Wise:2014ola,Wise:2014jva,Gresham:2017cvl,Gresham:2017zqi,Acevedo:2020avd}, consists of a dark matter fermion ($X$) coupled to a real scalar field ($\phi$) that provides an attractive force that binds the dark fermions together. It was recently shown by the authors that this dark matter model leads to interesting new signatures if the same scalar that binds the composite together has a small Yukawa coupling to Standard Model (SM) particles \cite{Acevedo:2020avd}. As we review below, sufficiently massive composites will be in a saturated state; the binding field inside the saturated composites takes on a classical value $\langle \phi \rangle \propto m_X$, where $m_X$ is the mass of the constituent dark matter fermion, here ranging from $\rm GeV$ - $\rm EeV$. For a wide range of couplings, these dark matter composites will source a large scalar field value, which implies a large corresponding scalar potential $V \sim \langle \phi \rangle$ inside the composite. Nuclei and other Standard Model particles coupled to $\phi$ will undergo accelerative processes at the composite boundary. It follows that the composite's Yukawa potential will cause nuclei (or other SM particles) to scatter, ionize, and undergo other dynamic processes at the boundary and inside the DM composite. 

As shown in \cite{Acevedo:2020avd}, the kinetic energy nuclei attain falling into large DM composites can result in Bremsstrahlung radiation and thermonuclear reactions. In particular, \cite{Acevedo:2020avd} explored how ionizing radiation can be observed at large neutrino observatories like IceCube and SNO+, as a large composite transits the detection volume of these experiments. In addition, there are important astrophysical consequences. For instance, the transit of these composites through massive white dwarf stars can ignite a Type-Ia supernovae. 

This work examines some new detection modes for large asymmetric composite dark matter states, including new searches at underground dark matter experiments, detailed computations for determining composite DM ignition of Type-Ia supernova, and the extent to which composites heat the Earth's interior. The outline of this paper is as follows: in Section~\ref{sec:sat-comp}, we discuss the basic properties of composite dark matter and review its cosmological synthesis. In Section~\ref{sec:nucaccel}, we introduce a Yukawa coupling to nucleons and show how this implies accelerative interactions near the composite boundary. In Section~\ref{sec:signatures}, we identify acceleration-based nuclear recoil signatures at underground dark matter search experiments, including atomic ionization through the Migdal effect. Section~\ref{sec:signatures} also investigates astrophysical signatures of nuclei accelerated by composite DM, including Earth heating and white dwarf explosions. In contrast to processes which accelerate nuclei in the presence of the DM composite's potential, Section \ref{sec:nucdmint} considers to what extent nuclei may scatter directly with individual constituents inside the DM composite. A detailed analysis shows that for most of the DM composite parameter space we consider, constituent scattering is negligible due to the highly degenerate composite interior. We conclude in Section~\ref{sec:concl}.

\section{Asymmetric Composite Dark Matter}
\label{sec:sat-comp}

Many details of asymmetric composite dark matter have been discussed in prior work $e.g.$ \cite{Wise:2014jva,Wise:2014ola,Hardy:2014mqa,Hardy:2015boa,Gresham:2017zqi,Gresham:2017cvl,Gresham:2018anj,Acevedo:2020avd}. Here we will focus on a model consisting of a Dirac fermion $X$, corresponding to the dark matter field, and a real scalar $\phi$, which mediates attractive self-interactions. The Lagrangian for this dark matter sector is
\begin{equation}
\mathcal{L}_D=\bar{X}(i\gamma^{\mu}\partial_{\mu}-m_{X})X+g_{X}\bar{X}\phi X+\frac{1}{2}m_{\phi}^{2}\phi^{2}+\frac{1}{2}\partial_\mu \phi \, \partial^\mu \phi,
\label{eq:lag1}
\end{equation} 
where the bound states we are interested are composed of $X$ (and not $\bar X$) particles, $i.e.$ the dark matter is asymmetric \cite{Petraki:2013wwa,Zurek:2013wia}.
An attractive interaction is a necessary ingredient to form bound states in the early universe, as we discuss shortly. Note that while we restrict our attention to the Lagrangian above for simplicity, much of the work we will do could be easily extended to the case that there is an additional repulsive vector coupling between $X$, or even pseudo-scalar and pseudo-vector interactions. Such interactions would tend to alter the formation and structure of the dark matter composites, see $e.g.$ \cite{Gresham:2018anj}. In what follows we shall also assume a zero temperature limit for our composites, but point out that the results outlined here can be generalized to a finite temperature using appropriate thermal distribution functions, see $e.g.$ \cite{Walecka:1995mi}.

The synthesis and physical properties of bound states with a low number of particles poses a complex physical problem. However, when the number of constituents is large we can apply relativistic mean field theory \cite{Gresham:2017zqi}, and approximate the binding field by a classical uniform value $\phi(x) \rightarrow \langle \phi \rangle$. In this limit, simple scaling relations in terms of the constituent number $N_X$ are recovered. The scalar field value inside the composite is determined by minimizing the energy density,
\begin{equation}
    \varepsilon = \frac{1}{2} m_\phi^2 \langle \phi \rangle^2 + \frac{1}{\pi} \int_0^{p_F} dp \ p^2 \left(p^2 + m_*^2\right)^{1/2}.
    \label{eq:sat-comp-edensity}
\end{equation}
In this expression, $m_* = m_X - g_X \langle \phi \rangle$ is the fermion effective mass inside the composite, which accounts for the $X$ self-interactions. The upper integration limit is the Fermi momentum $p_F$, which implicitly depends on the scalar field expectation value in the composite, $\langle \phi \rangle$, via the chemical potential $\mu = (p_F^2+m_*^2)^{1/2}$. In order to determine $\langle \phi \rangle$ by minimizing Eq.~\eqref{eq:sat-comp-edensity}, the chemical potential must in turn be related to the energy density via $\mu = \varepsilon / n_X$, where $n_X = p_F^3/3\pi^2$ is the constituent number density. This is equivalent to requiring a vanishing pressure, since $p = -(\partial E / \partial V)_{N_X} = \mu \varepsilon - n_X$. In the limit $m_* \ll p_F$, simple scaling relations for the effective mass and chemical potential are recovered,
\begin{equation}
    m_* \simeq (6 \pi^2)^{1/2} \left(\frac{m_\phi}{g_X}\right),
    \label{eq:sat-comp-meff}
\end{equation}
\begin{equation}
    \mu \simeq p_F \simeq (6 \pi^2)^{1/2} \left(\frac{m_\phi}{g_X}\right)^{1/2} \left(\frac{m_X}{g_X}\right)^{1/2}.
    \label{eq:sat-comp-chem}
\end{equation}
It can be seen from the above expressions that $\mu \simeq p_F \gg m_*$, and so the constituents are effectively relativistic while in the bound state. This will be especially relevant for computing nucleus-$X$ scattering, see Section~\ref{sec:nucdmint2}. The binding energy per constituent is set by the difference $m_X - \mu$, and related to the total composite mass via
\begin{equation}
    M_X = N_X m_X - N_X (m_X - \mu) = N_X \mu
\end{equation}

Using a standard ``liquid drop'' model, the total composite mass is made up of a bulk and surface contribution $M_X = N_X \bar{m}_X +  N_X^{2/3} \epsilon_{\rm surf}$, where $\bar{m}_X$ is the amount of composite mass per constituent, which accounts for their binding energy, and $\epsilon_{\rm surf}$ is a coefficient associated with the decrease of interactions near the composite boundary. This relation implies that the chemical potential scales as $\mu = \bar{m}_X + N_X^{-1/3} \epsilon_{\rm surf} \simeq \bar{m}_X$ in the limit of large $N_X$. Thus, from here onward we will take the composite mass as $M_X \simeq N_X \bar{m}_X$, and because the constituents are relativistic, the mass they contribute to the composite is simply $\bar{m}_X \simeq \mu $, which for relativistic constituents will be their Fermi momentum, $\bar{m}_X \simeq (6 \pi^2 m_\phi m_X)^{1/2}/g_X$ as given by Eq.~\eqref{eq:sat-comp-chem}. The composite radius is consequently,
\begin{equation}
    R_X = \left(\frac{9 \pi N_{X}}{4 \bar{m}_{X}^{3}}\right)^{1/3},
    \label{eq:sat-comp-radius}
\end{equation}
where the scaling indicates that the composite volume is proportional to the composite mass. Therefore, the addition of an extra particle simply enlarges the composite so that the number density remains constant. This saturation number density is 
\begin{equation}
    n_X = \frac{\bar{m}_X^3}{3 \pi^2} \simeq 10^{41} \ {\rm cm^{-3}} \ \left(\frac{\bar{m}_X}{5 \ \gev}\right)^3.
\end{equation}

Composites with a number density exceeding the above expression will be well-described by the mass and radius relations given above. Of course it is often preferable to quantify the minimum number of dark matter constituents which render the above approximations to be accurate. As discussed extensively in \cite{Gresham:2017zqi}, the saturation regime can be defined as the limit where the composite length scale $R_X$ becomes comparable to the mediator range $m_\phi^{-1}$, implying that a composite state must contain a minimum
\begin{equation}
    N_X \simeq \left(\frac{\bar{m}_X}{m_\phi}\right)^3 \simeq 10^{10} \ \left(\frac{\alpha_X}{0.3}\right)^\frac{3}{4} \left(\frac{m_X}{\rm TeV}\right)^\frac{3}{2} \left(\frac{m_\phi}{\mev}\right)^{-2} 
    \label{eq:sat-com-nmin}
\end{equation}
of dark matter constituents for it to be saturated. Shortly we will see that by the time that the cosmological synthesis of these composites ceases, the typical particle numbers $N_X$ are well above this saturation threshold.

The cosmological formation of large asymmetric composite states has been explored in previous works \cite{Wise:2014jva,Hardy:2014mqa,Gresham:2017cvl,Acevedo:2020avd}. In the case of the composite dark sector indicated by Eq.~\eqref{eq:lag1}, composite synthesis occurs in the early universe when the dissociation of two-body bound states by $\phi$-scattering becomes inefficient as the universe cools. Formation of two-body bound states will occur provided  that \cite{Wise:2014jva}
\begin{equation}
    \alpha_X^2 m_X \gtrsim m_\phi 
\end{equation}
and 
\begin{equation}
    \alpha_X \gtrsim 0.3 \left( \frac{m_X}{10^7\,{\rm GeV}} \right)^{\frac{2}{5}} \left(\frac{\zeta}{10^{-6} }\right)^{\frac{1}{5}}
\end{equation}
with $\alpha_X = g_X^2 / 4\pi$. The parameter $\zeta \ll 1$ is a dilution factor that ensures the observed dark matter relic abundance is attained at the end of synthesis. We specify its origin a few lines below. Once two-body bound states are formed, synthesis of larger composites proceeds via inelastic fusion processes, with cooling of the final composite usually occurring via $\phi$ emission. In the strong binding regime $\bar{m}_X \ll m_X$, the composites will build up in size via dark fusion reactions of the form $^{N_X} \! X$ + $^{N_X} \! X$ $\rightarrow$ $^{2N_X} \! X$ + $\phi$, until their rate drops below the expansion rate of the universe. For masses $m_{X} \gtrsim \rm TeV$, this results in an overabundance of dark matter. However, the correct relic abundance can be recovered assuming the excess is subsequently depleted by the decay of a metastable field \cite{Bramante:2017obj}, as was recently shown in \cite{Acevedo:2020avd}. Such a late-time relic density depletion via decaying fields is present in many models of high-scale baryogenesis \cite{Affleck:1984fy,Dine:1995kz}. The amount of asymmetric dark matter (and baryons) that are depleted, is given by $\Omega_{X}^{\rm dep} = \Omega_{X} \zeta$, where $\zeta = s_{\rm after} / s_{\rm before}$ is the entropy ratio before and after the field decays. Conversely, the abundance of dark matter constituent particles $X$ after freeze-out is in excess of the standard value by a factor of $\zeta^{-1}$, leading to the formation of large composite states. The number of dark matter of particles contained in a composite formed by this process is \cite{Acevedo:2020avd},
\begin{align}
    \label{eq:cosmo1}
    &N_{c}  = \left(\frac{20 \sqrt{g^*_{ca}} T_{r} T_{ca}^{3/2} M_{pl}}{\bar m_X^{7/2} \zeta} \right)^{6/5} \\
    &\simeq 10^{27} \left(\frac{g^*_{ca}}{10^2} \right)^{\frac{3}{5}} \left(\frac{T_{ca}}{10^5\,{\rm GeV}} \right)^{\frac{9}{5}}  \left(\frac{5\,{\rm GeV}}{\bar m_X} \right)^{\frac{21}{5}}\left(\frac{10^{-6}}{\zeta} \right)^{\frac{6}{5}},\nonumber
\end{align}
where $T_r \simeq 0.8~{\rm eV}$ is the temperature of matter-radiation equality, $T_{ca} \simeq m_X/10$ is the $X$ freeze-out temperature, $g_{ca}^*$ is the number of relativistic field degrees of freedom at freeze-out, and $M_{pl}$ is the reduced Planck mass.
From this we see that in the regime in which consistent cosmological formation can be obtained for dark fermion masses up to and even exceeding $10^{10} \ \rm GeV$. The resulting number of dark matter particles contained per composite is well above the saturation threshold, Eq.~\eqref{eq:sat-com-nmin}, and leads to composite masses approximately ranging over $10^{10}$ - $10^{42} \ \gev$, where even larger masses may be obtained for a smaller dilution parameter $\zeta$. Thus, the binding field in the composite interior is $\langle \phi \rangle \propto m_X$, and potentially results in highly-energetic signatures, as explored in \cite{Acevedo:2020avd}. However, as we will see in the following sections, even if the composites are comparatively less massive and $\langle \phi \rangle$ is correspondingly smaller ($i.e.$ for $\zeta =1$), even a small coupling between $\phi$ and Standard Model nucleons can result in detectable nuclear scattering and ionization processes at underground dark matter experiments.

\section{Nuclear Coupling and Acceleration}
\label{sec:nucaccel}

So far we have discussed the properties of saturated composite states without assuming any coupling of this dark sector to Standard Model particles. We now introduce a coupling between the binding field $\phi$ and SM nucleons by adding a Yukawa interaction,
\begin{equation}
    \mathcal{L}_{n}=g_{n}\bar{n} \phi n
\end{equation}
to Eq.~\eqref{eq:lag1}, where $g_{n}$ is the Yukawa coupling between $\phi$ and Standard Model nucleons. Note that $g_n$ can be either positive or negative, resulting in either a repulsive or attractive Yukawa potential for nucleons sourced by the DM composite. The binding field $\phi(x)\simeq \langle \phi \rangle$ is classical and effectively uniform inside a large composite, and its value is well-approximated by
\begin{equation}
\langle \phi \rangle \simeq \frac{m_{X}}{g_{X}}, \ \ \ r < R_{X},
\end{equation}
so long as $m_* \ll p_F$, $cf.$ Eq.~\eqref{eq:sat-comp-meff}. On the other hand, because of boundary conditions, the field must rapidly decay outside the composite state according to,
\begin{equation}
\phi(r)=\langle \phi \rangle \ e^{-m_{\phi}(r-R_{X})} \left(\frac{R_{X}}{r}\right), \ \ \ r \geqslant R_{X}.
\label{eq:accelphi}
\end{equation}
Nuclei coupled to this field $\phi$ will then have an effective mass smaller or larger than their vacuum mass in the composite interior, for an attractive and repulsive potential, respectively. Energy conservation imposes that their momenta must change as they enter the boundary of the composite state (and similarly as they exit) according to,
\begin{equation}
 p^{2}+m_{N}^{2}=p'^{2}+\left(m_{N}-\langle \varphi \rangle \right)^{2},
 \label{eq:accelsr1}
\end{equation} 
where $p$ and $p'$ are the momenta of a nucleus before and after entering the composite respectively, $m_{N} = A m_{n}$ is the nuclear mass, and we have defined $\langle \varphi \rangle=A g_{n} \langle \phi \rangle$. We note again that the above implies a decrease or increase in the mass of the nucleus, depending on the sign of $g_n$. 

Before continuing, for the sake of clarity we summarize four interaction regimes for large DM composites with a Yukawa coupling to nuclei.
\begin{enumerate}
    \item $\langle \varphi \rangle \lesssim m_N$ and $g_n > 0$. In this case there will be energetic interactions between nuclei accelerated into the composite interior. We have recently studied signatures of the resulting Bremsstrahlung and fusion processes in \cite{Acevedo:2020avd}.
    \item $\langle \varphi \rangle \ll m_N$ and either $g_n >0$ or $g_n < 0$. Low energy attractive or repulsive interactions will result in a soft nuclear recoil at the boundary of the composite, yielding scintillation and Migdal ionization at underground experiments as we show for the first time in this work. Quite recently, repulsive DM composites were also studied in the context of mineralogical detection \cite{Ebadi:2021cte}.
    \item $\langle \varphi \rangle \gg m_N$ and either $g_n > 0$ or $g_n < 0$. For a large enough (and hence relativistic) attractive or repulsive composite potential, the DM composite boundary will form a repulsive barrier for incoming nuclei \cite{Greiner:1990tz}, as discussed in \cite{Acevedo:2020avd}. Most of the nuclear recoil work presented here will generalize easily to this case.
    \item In addition to interactions between the dark matter composite's scalar potential $\langle \varphi \rangle$ and nuclei, we also consider nucleus-$X$ (that is nucleus-constituent) scattering interactions. These can occur as scattering interactions between nuclei and single DM constituents, or take the form of collective excitations of multiple composite constituents. As detailed in Section \ref{sec:nucdmint} (see also Appendix~\ref{sec:app-b}), these interactions are highly suppressed relative to interactions with the composite potential, since $X$ constituent particles in saturated composites are highly degenerate.
\end{enumerate}

\begin{figure}
     \centering
     \centerline{\includegraphics[width=1.1\textwidth]{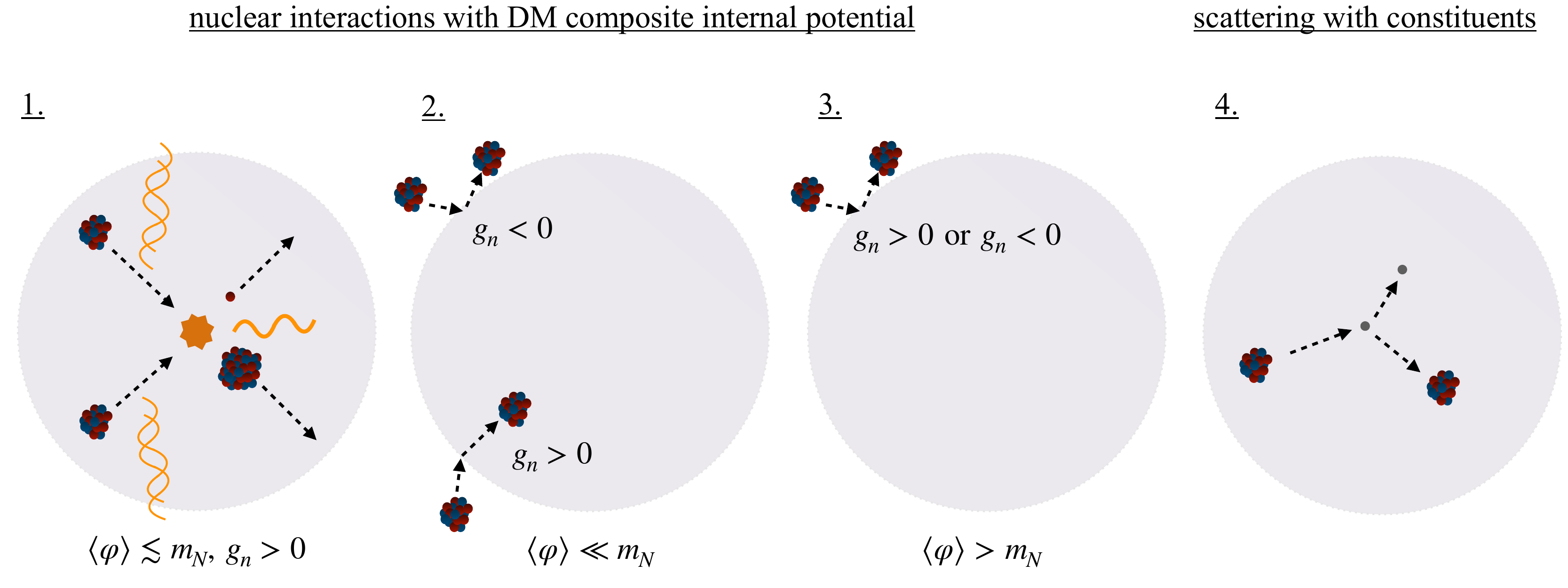}}
    
     \caption{Different interactions are shown between a large DM composite and nuclei, for nucleons that have a Yukawa coupling $g_n$ to the scalar field $\phi$ that binds the composite constituents $X$ together. For the first three processes, the size of the Yukawa potential for nuclei inside the DM composite, $\langle \varphi \rangle=A g_{n} \langle \phi \rangle$, as compared to the nuclear mass $m_N$, determines the characteristics of nuclear interactions with the composite's scalar potential. The sign of the Yukawa coupling $g_n$ determines whether the potential is attractive or repulsive. {\em 1.}~The first panel shows that for a sizable, attractive, but non-relativistic Yukawa potential, nuclei emit Bremmstrahlung radiation and may fuse in the composite interior, as studied in \cite{Acevedo:2020avd}. {\em 2.}~The second panel shows that for a small composite potential $\langle \varphi \rangle \ll m_N$, nuclei will recoil at the boundary of the composite, either accelerating or decelerating, depending on whether the potential is attractive or repulsive. Note that in the case of a repulsive Yukawa potential, the extent to which the nucleus enters the composite will depend on its initial kinetic energy as compared to $\langle \varphi \rangle$. {\em 3.}~The third panel shows that for a relativistic internal potential, there will be a large potential barrier that reflects nuclei \cite{Acevedo:2020avd,Greiner:1990tz}. {\em 4.}~The fourth panel depicts nuclear scattering with DM constituents $X$. As detailed in Section \ref{sec:nucdmint2}, this sort of scattering interaction will be very suppressed since the $X$ particles in the composite are highly degenerate.}
    \label{fig:DMschem}
\end{figure}

Figure \ref{fig:DMschem} shows a schematic of the different recoil and high energy processes that arise from a Yukawa coupling between the binding field for composites ($\phi$) and SM nucleons. In what follows we will focus on the case that $g_n$ is positive and the potential is attractive. For the low-energy scattering processes we will study in Section \ref{sec:migdal}, the generalization to $g_n$ negative is straightforward, since the nuclear recoil spectrum implied for an attractive and repulsive potential are the same. 

To quantify the kinetic energy change of nuclei at the boundary of the composite, we first note that the field $\langle \varphi \rangle$ couples directly to the nuclear mass since it is a Lorentz scalar. When $\langle \varphi \rangle \ll m_{N}$, we neglect the $\sim \mathcal{O}(\langle \varphi \rangle^{2})$ term and obtain the kinetic energy change in the non-relativistic limit,
\begin{equation}
\Delta{E} \simeq \left(\frac{p'^{2}}{2m_{N}}-\frac{p^{2}}{2m_{N}}\right) \simeq A g_{n} \left(\frac{m_{X}}{g_{X}}\right).
\label{eq:accelsr3}
\end{equation}
In this limit, the kinetic energies of nuclei can lead to nuclear fusion and bremmstrahlung we investigated in \cite{Acevedo:2020avd}. Eq.~\eqref{eq:accelsr3} above illustrates the central result: the energy difference per SM nucleon is given by $g_n m_{X}/g_{X}$ for large composites, where the constituent mass $m_X$ can span several orders of magnitude in straightforward cosmologies, ranging from sub-GeV up to EeV. Such heavy constituents are accommodated in a cosmology that includes a stage of relic abundance dilution \cite{Acevedo:2020avd}, $cf.$ Eq.~\eqref{eq:cosmo1}. Therefore, even for tiny coupling values, nuclei can gain or lose substantial kinetic energy as they cross the composite boundary. In addition, the composite states considered have large sizes compared to the atomic separation at densities $\rho_* \gtrsim 1 \ \rm g \ cm ^{-3}$, and so they collect a large number of nuclei in their interior. Thus, composites with this simple renormalizable coupling to SM nucleons act as microscopic nuclear accelerators as they traverse matter.  

A number of processes can occur between nuclei and DM composites, depending on the Yukawa coupling $g_n$, as this quantity determines the dynamics of SM matter as it crosses the composite interior. For nuclear kinetic energy shifts $\Delta E \lesssim 100 \ \rm eV$, a small fraction of the entering atoms will be ionized via collisional processes and the Migdal effect \cite{Migdal:1977bq,Landau:1991wop}, leading to the emission of electromagnetic radiation. In Section~\ref{sec:migdal}, we demonstrate that the Migdal effect alone permits liquid noble element experiments like \acro{XENON-1T} to probe nucleon-scalar Yukawa couplings well below existing constraints based on stellar cooling arguments. For $\Delta E \gtrsim 100 \ \rm eV$, low-Z atomic nuclei will be fully ionized, leading to substantial thermal Bremsstrahlung, as explored in \cite{Acevedo:2020avd}. There, it was shown that this radiation induced during nuclear transit through the composite leads to a massive energy release detectable by neutrino observatories like IceCube and SNO+, for composite dark matter in the mass range $10^{21} \ \gev \lesssim M_X \lesssim 10^{25} \ \gev$. At temperatures $T \gtrsim 100 \ \kev$ - $1 \ \mev$, thermonuclear reactions may proceed, however the specific reactions and their rate will depend on the composition, density and temperature of the given medium. Ref.~\cite{Acevedo:2020avd} previously showed that the heat deposition is adequate to trigger a thermonuclear runaway in white dwarfs. In Section \ref{sec:IaSNe}, we expand our discussion of Type-Ia supernova ignition, and go on to explore implications for planetary heating in Section~\ref{sec:earthflow}. At much higher energies, the non-relativistic approximation made here breaks down. This can be seen from Eq.~\eqref{eq:accelsr1}: if $\langle \varphi \rangle \gtrsim m_{N}$, then necessarily $p' < p$, implying that this potential eventually becomes repulsive in this limit \cite{Greiner:1990tz}.  We have left a detailed treatment of this case to future work -- however we would point out that there are bounds on such high values of $\langle \varphi \rangle$, because of stellar cooling limits considerations, which at present limit the coupling to values $g_n \lesssim 10^{-10}$ - $10^{-12}$ over the mediator mass range $m_\phi \sim {\rm eV}$ - $\rm GeV$.

To conclude this Section, we discuss the timescale over which nuclei are accelerated when entering the composite's Yukawa potential. As we show below, this timescale is short compared to the timescale over which the nucleus crosses the DM composite, since the composite radii considered here are much larger than the binding force range, and so nuclei are accelerated over a short distance to substantial energies. First, we consider the dynamical expression for the acceleration time

\begin{equation}
\tau_{\rm accel}=\left(\frac{m_{N}}{2}\right)^{1/2}\int_{\infty}^{R_{X}} \left(\varepsilon+\varphi(r)\right)^{-1/2} \ dr,
\label{eq:accel1}
\end{equation}
where $\varepsilon$ denotes the energy of the nucleus in the composite rest frame and $\varphi(r)= A g_{n} \phi(r)$, with $\phi(r)$ given by Eq.~\eqref{eq:accelphi}. The initial energy $\varepsilon$ can be expressed in terms of the energy in the laboratory frame $\varepsilon_0$ via $\varepsilon = \varepsilon_{0}+\mathbf{p}\cdot \mathbf{v}_{X}+m_{N} v_{X}^{2}/2$, with $\mathbf{v}_{X}$ being the velocity of the composite in such frame. For the phenomenology considered in this work, $\varepsilon_{0} \ll m_{N} v_{X}^{2}/2$ and the initial (pre-composite crossing) energy of nuclei in the composite rest frame is well approximated by $\varepsilon \simeq m_{N} v_{X}^{2}/2$. Because the field gradient decays exponentially outside the composite volume, nuclei accelerate over a short distance near the composite surface. In terms of the dimensionless variable $\chi=r/R_{X}$, this length scale is $\chi \simeq (m_{\phi}R_{X})^{-1}$. The acceleration timescale is then
\begin{equation}
\tau_{\rm accel}=R_X \left(\frac{m_{N}}{2\varepsilon}\right)^{1/2} \int_{1+(m_{\phi}R_{X})^{-1}}^{1} \left(1+\left(\frac{\varphi}{\varepsilon}\right)\frac{e^{-m_{\phi}R_{X}(\chi-1)}}{\chi}\right)^{-1/2} \ d\chi,
\end{equation}
which can be solved numerically. However, for the large composite masses we consider, it is always the case that $(R_{X} m_{\phi})^{-1} \ll 1$. Therefore, we can estimate the timescale above by approximating $\chi \simeq 1$ and $d\chi \simeq (m_{\phi}R_{X})^{-1}$, yielding
\begin{equation}
\tau_{\rm accel}\simeq \frac{1}{m_\phi} \left(\frac{1}{v_X^2 + v_N^2}\right)^{1/2} \simeq 6.5 \times 10^{-19} \ {\rm s} \ \left(\frac{m_{\phi}}{\mev}\right)^{-1} \left(\frac{v_X}{10^{-3}}\right)^{-1},
\label{eq:acceltime}
\end{equation}
where $v_N = (2\langle \varphi \rangle/m_N)^{1/2}$ is the final nuclear velocity in the composite rest frame. In the rightmost expression, we show the scaling when $v_N \ll v_X$, which is the case for all phenomena considered in this work.\footnote{We note that in the opposite limit $v_X \ll v_N$ a finite acceleration timescale is also recovered
\begin{equation}
\tau_{\rm accel} \simeq m_\phi^{-1} \left(\frac{2\langle \varphi \rangle}{m_N} \right)^{-1/2} \simeq 10^{-18} \ {\rm s} \ \left(\frac{m_N}{10 \ \gev} \right)^{\frac{1}{2}} \left(\frac{m_\phi}{\mev} \right)^{-1} \left(\frac{\langle \varphi \rangle}{\mev}\right)^{-\frac{1}{2}}.
\end{equation}}
Having now derived the acceleration timescale for nuclei entering the DM composite, we turn to associated signatures.

\section{Nuclear Acceleration Signatures of Composite DM}
\label{sec:signatures}

Standard model particles accelerated across the boundary of dark matter composites provide new means for dark matter detection. These include nuclei recoiling across the composite boundary at underground experiments, and the associated Migdal effect, along with astrophysical signatures of composites heating planets and igniting white dwarfs. In this Section we detail the detection of accelerating dark matter composites using underground dark matter search experiments, along with white dwarf and terrestrial observations.

\subsection{Direct Searches via Nuclear Recoil and the Migdal Effect at Low Energies}
\label{sec:migdal}

Composite dark matter crossing the volume of direct detection experiments can accelerate atomic nuclei over a short timescale, thereby ionizing and exciting atoms with the sudden nuclear recoil induced by the DM composite's potential. We first turn to the ionization of atoms at the DM composite boundary. The non-adiabatic response of electrons to an impulsive nuclear motion is called the Migdal effect \cite{Migdal:1977bq,Landau:1991wop}. In essence, the Migdal formalism relies on a sufficiently rapid change in nuclear momentum, so that the perturbed electron wavefunctions can be modeled by applying a straightforward momentum boost. Therefore, to apply the Migdal formalism to large DM composites we will require that the nuclear recoil interaction time is short compared to both the electron orbital period $\tau_{e^-} \sim (10 \ {\rm eV})^{-1}\simeq 6.5 \times 10^{-17} \ \rm s$ and the ratio $R_a / v_N$, where $R_a$ is the atomic radius and $v_N$ is the final nuclear velocity. Before the interaction the atomic nucleus can be assumed stationary, and the electron cloud is characterized by its ground state wavefunction $|\psi\rangle$. After the interaction has occurred, the nucleus moves with speed $v_N$. In the rest frame of the recoiling nucleus, if the interaction occurred fast enough, the electrons initially have the same coordinates as when the nucleus was stationary, but their momenta are boosted by $q = m_e v_N$. The perturbed electron cloud wavefunction is subsequently expressed as $|\psi'\rangle \simeq \exp\left(-i \sum_a \mathbf{q} \cdot \mathbf{r}_a\right) |\psi \rangle$, where $r_a$ are the position operators of the electrons in this new rest frame, with $a=1,...,Z$. This yields a finite transition amplitude to excited and free states, even for the electrons occupying inner orbitals. 

In practice, noble elements used in direct dark matter searches have large atomic radii of order $R_a \simeq 10^{-8} \ \rm cm$, so $R_a / v_N \gg \tau_{e^-}$. Comparing then the electron orbital period to Eq.~\eqref{eq:acceltime}, we see that the acceleration occurs fast enough for this approximation to be valid, so long as the composite is moving sufficiently fast or the potential is sufficiently short-ranged. In this Section we will find that Migdal electrons produced by the composite transit yield excellent detection prospects for liquid noble element experiments like \acro{XENON-1T}.

There have been several proposals for dark matter searches using the Migdal effect \cite{Vergados:2004bm,Moustakidis:2005gx,Ejiri:2005aj,Bernabei:2007jz,Ibe:2017yqa,Dolan:2017xbu,Essig:2019xkx,Baxter:2019pnz,GrillidiCortona:2020owp,Liu:2020pat,Flambaum:2020xxo,Knapen:2020aky,Bell:2021zkr}, especially in the sub-GeV mass range. As a proof-of-concept for DM composite detection, we will focus here on \acro{XENON-1T} and its first dark matter search  \cite{Aprile:2017iyp}. As we show below, the extremely low electron background of this experiment, combined with the high mass number of xenon, allows for a sensitivity to dark matter composite-nucleon couplings well below existing constraints. Detection prospects using the Migdal effect have been conducted previously in \cite{Dolan:2017xbu}. Here, because of the nature of the Migdal effect as applied to DM composites, the recoil energy spectrum will be different because all atomic nuclei are accelerated to approximately the same kinetic energy along the composite's path. 

For a large DM composite, the Migdal ionization differential event rate per unit of exposure is set by
\begin{equation}
    \frac{dR}{dE_R} = \frac{\rho_X}{m_N M_X} \int_{v > v_X^{(\rm min)}} \frac{d\sigma}{dE_R} \ v \ g(v) \ dv
    \label{eq:Migdaleventrate-general}
\end{equation}
where $m_N \simeq 130 \ \gev$ is the mass of a xenon atom, and $g(v)$ is the dark matter velocity distribution in the laboratory frame. In our analysis we assume a local dark matter density $\rho_X \simeq 0.4 \ \rm GeV \ cm^{-3}$ \cite{Iocco:2011jz} and use a flux-normalized velocity distribution outlined in \cite{Acevedo:2020gro},

\begin{equation}
    f(\mathbf{v}) = \mathcal{N}^{-1} (v^2-v_e^2)^{3/2} \exp \left(\frac{\tilde{v}^2}{v_0^2}\right) \Theta(v-v_e) \Theta(v_{eg}-\tilde{v}),
    \label{eq:Migdalvdist}
\end{equation}
where $v=|\mathbf{v}|$, $v_e \simeq 11.2 \ \rm km \ s^{-1}$ is the Earth's escape velocity, $v_{eg} \simeq 528 \ \rm km \ s^{-1}$ is the galactic escape velocity, and $v_0 \simeq 220 \ \rm km \ s^{-1}$ is the velocity dispersion \cite{Read:2014qva,Pato:2015dua}. The variable $\tilde{v}^2 = v^2 - v_e^2 +v_{rf}^2 + 2v_{rf} \sqrt{v^2-v_e^2} \cos \varphi$ accounts for the relative velocity between the solar system with respect to the galactic frame, with $v_{rf} \simeq 230 \ \rm km \ s^{-1}$ and $\varphi$ being the angle between such relative velocity and the dark matter vector. The constant $\mathcal{N}^{-1}$ is a normalization factor that enforces $\int_0^{\infty} \int_{0}^{\pi} f(\mathbf{v}) \ dv \ d(\cos\varphi) = 1$. The distribution $g(v)$ in Eq.~\eqref{eq:Migdaleventrate-general} is then given by the integration over $\varphi$ of Eq.~\eqref{eq:Migdalvdist}, $i.e.$ $g(v) = \int_{0}^{\pi} f(\mathbf{v}) d(\cos\varphi)$.

We express the differential cross section per unit of nuclear recoil energy as the geometric cross section of the composite, multiplied by a delta function enforcing the fact that all nuclei along the composite path will be accelerated to the same energy

\begin{equation}
   \frac{d\sigma}{dE_R} \simeq 2 \pi R_X^2 \  \delta \! \left(E_R - E_R^0\right)
   \label{eq:X1TdsigdER}
\end{equation}
where $E_R^0 = A g_n m_X / g_X$ ($A\simeq 130$) denotes the kinetic energy to which nuclei are accelerated as they cross the composite boundary, $cf.$ Eq.~\eqref{eq:accelsr3}. Note that there is a factor of two in this expression that accounts for nuclei accelerating both as they approach and exit the composite boundary.

The minimum integration speed in Eq.~\eqref{eq:Migdaleventrate-general}, unlike other analyses, is now set by the minimum kinetic energy that the composite must have to penetrate the experiment's overburden, and still reach the detection volume with a sufficient speed for the Migdal approximation to remain valid. Depending on the coupling and mediator mass, radiation or conduction losses will dominate the stopping power. For couplings sufficiently high so that temperatures $T \gtrsim 100 \ \rm eV$ are reached in the composite interior, matter will be completely ionized but optically thin to photons, resulting in an energy loss in the form of thermal bremsstrahlung \cite{Acevedo:2020avd}. At lower couplings, heat conduction losses will dominate, which we estimate by taking a crust thermal conductivity $\sim 1 \ {\rm W \ m^{-1} \ K^{-1}} \simeq 10^7 \ \rm GeV \ s^{-1}$ \cite{beardsmore2001crustal} and applying Fourier's law with a thermal gradient $\nabla T \simeq T m_\phi$. Both of these energy loss channels imply that composites must have a minimum velocity of order $v_X^{\rm (heat)} \gtrsim 10^{-4}$ - $10^{-3}$, and so an $\sim \mathcal{O}(1)$ fraction of the flux will have enough kinetic energy to penetrate the overburden. On the other hand, for the Migdal approach to be valid, we require the acceleration timescale $\tau_{\rm accel}$, given by Eq.~\eqref{eq:accel1}, to be short enough compared to the electron orbital period, of order $\tau_{e^-} \sim 10^{-17} \ \rm s$. Thus, setting $\tau_{\rm accel} = \tau_{e^-}$ in Eq.~\eqref{eq:accel1}, we obtain the minimum speed

\begin{equation}
    v_X^{\rm (mig)}\simeq \frac{1}{m_\phi \tau_{e^-}} \simeq 10^{-5} \left(\frac{\mev}{m_\phi}\right),
\end{equation}
where we have neglected $v_N$ in Eq.~\eqref{eq:accel1} compared to $(m_\phi \tau_{e^-})^{-1}$, which is the case for the entire parameter space considered here. 

The differential ionization rate will be given by Eq.~\eqref{eq:Migdaleventrate-general}, multiplied by the probability of electron emission from a given energy level \cite{Dolan:2017xbu},

\begin{equation}
    \frac{dR_{\rm ion}}{dE_R dE_e} =  \frac{dR}{dE_R} \times \left(\frac{1}{2 \pi} \sum_{n,l} \frac{dp_{n,l \rightarrow E_e}}{dE_e} \bigg|_{q} \right).
    \label{eq:drion}
\end{equation}
in the above equation, $E_e$ is the final kinetic energy of the ionized electron and $\frac{dp_{n,l \rightarrow E_e}}{dE_e}|_{q}$ is the differential probability, for a given momentum change $q = m_e v_N = (2m_e^2 E_R^0/m_N)^{1/2}$, for an electron initially at a level $(n,l)$ to be ionized with a final kinetic energy $E_e$. This set of probabilities was numerically computed for xenon atoms in \cite{Ibe:2017yqa}; we use the results of that study to evaluate the integral in Eq.~\ref{eq:drion} (Table II of \cite{Ibe:2017yqa}). The total energy deposition is given by $E_{em} = E_{nl} + E_e$, where $E_{nl}$ is the initial binding energy of the electron. We remark that while $E_e$ is typically $\sim \mathcal{O}(\rm eV)$, the ionization energies $E_{nl}$ are $\sim \mathcal{O}(\kev)$, and therefore dominant.

Due to the very narrow nuclear recoil spectrum, $cf.$ Eq.~\eqref{eq:X1TdsigdER}, integration over the nuclear recoil energy can be performed analytically. We can also integrate over the electronic energies and obtain

\begin{equation}
    R_{\rm ion} =  \frac{4 \pi R_X^2 n_X}{m_N} \times \left( \int_{v > v_X^{(\rm min)}} \ v \ g(v) \ dv \right) \times \left(\frac{1}{2 \pi} \sum_{n,l} \int dE_e \ \varepsilon(E_{em}) \ \frac{dp_{n,l \rightarrow E_e}}{dE_e} \bigg|_{q} \right).
\end{equation}

We have explicitly included here \acro{XENON-1T}'s detection efficiency for a given electromagnetic energy deposition $\varepsilon(E_{em})$ \cite{Dolan:2017xbu}. For \acro{XENON-1T}'s first run, a total exposure of $98 \ \rm kg \ yr$ was achieved \cite{Aprile:2017iyp}. Furthermore, the integral in the above equation yields a factor of $q^2$, allowing us to express the total number of Migdal electron events at \acro{XENON-1T} in terms of composite parameters, $cf.$ Eqs.~\eqref{eq:sat-comp-radius} and \eqref{eq:accelsr3},

\begin{equation}
    N_{\rm ion} \simeq 10^8 \left(\frac{m_X}{10^3 \ \gev}\right)^{-\frac{2}{5}} \left(\frac{m_\phi}{10^{-3} \ \gev}\right)^{-\frac{4}{5}} \left(\frac{g_n}{10^{-10}}\right) \left(\frac{\alpha_X}{0.3}\right)^{-\frac{1}{10}}.
    \label{eq:Migdaliontot}
\end{equation}
It is also relevant to compute the average number of electronic recoils produced by a single composite transiting the detection volume. This is given by the ionization probabilities, the detection efficiency, and the flux of xenon atoms through the composite,

\begin{align}
    N_{\rm transit} \simeq & \left(2 \pi R_X^2 n_{\rm Xe} L_{\rm det}\right) \times \left(\frac{1}{2 \pi} \sum_{n,l} \int dE_e \ \varepsilon(E_{em}) \ \frac{dp_{n,l \rightarrow E_e}}{dE_e} \bigg|_{q}\right) \nonumber \\ & \simeq 5 \times 10^{16} \left(\frac{R_X}{\rm nm}\right)^2 \left(\frac{m_X}{\rm TeV}\right) \left(\frac{g_n}{10^{-10}}\right) \left(\frac{\alpha_X}{0.3}\right)^{-\frac{1}{2}}.
    \label{eq:Migdalionsingle}
\end{align}
In the above expression, $L_{\rm det} \simeq 100 \ \rm cm$ is the length scale of \acro{XENON-1T}'s detection volume, $n_{\rm Xe} \simeq 10^{22} \ \rm cm^{-3}$ is the number density of targets, and we assume the composite is moving fast enough for the Migdal approximation to be valid.  Eq.~\eqref{eq:Migdalionsingle} demonstrates that even for tiny couplings, a composite crossing \acro{XENON-1T}'s volume could be observed from the large ionization track it creates.

Finally, we must consider the maximum composite mass that can be probed by \acro{XENON-1T} given the total dark matter flux passing through the experiment. An accurate modelling of the experiment's geometry is out of the scope of this work (see, $e.g.$, \cite{Bramante:2018qbc,Bramante:2018tos,Clark:2020mna,Bhoonah:2020fys,Acevedo:2021tbl}), however as a first estimate we use the flux through a spherical detector with a radius of order $R_{\rm det} \sim 50 \ \rm cm$, yielding $\Phi_X \simeq 2\pi R_{\rm det}^2 \langle v_X \rangle (\rho_X/M_X) \simeq 10^{-4} \ {\rm s^{-1}}  \left( M_X/10^{15} \ \gev \right)^{-1}$. The maximum mass to which the experiment is sensitive is then estimated by requiring $\Phi_X t \simeq 1$, where $t$ is the running time. \acro{XENON-1T}'s first dark matter search had a total of 34.2 live days \cite{Aprile:2017iyp}, resulting in a mass limit of $M_X^{\rm (max)} \simeq 2 \times 10^{17} \ \gev$.

\begin{figure}
     \centering
     \centerline{\includegraphics[width=1.15\textwidth]{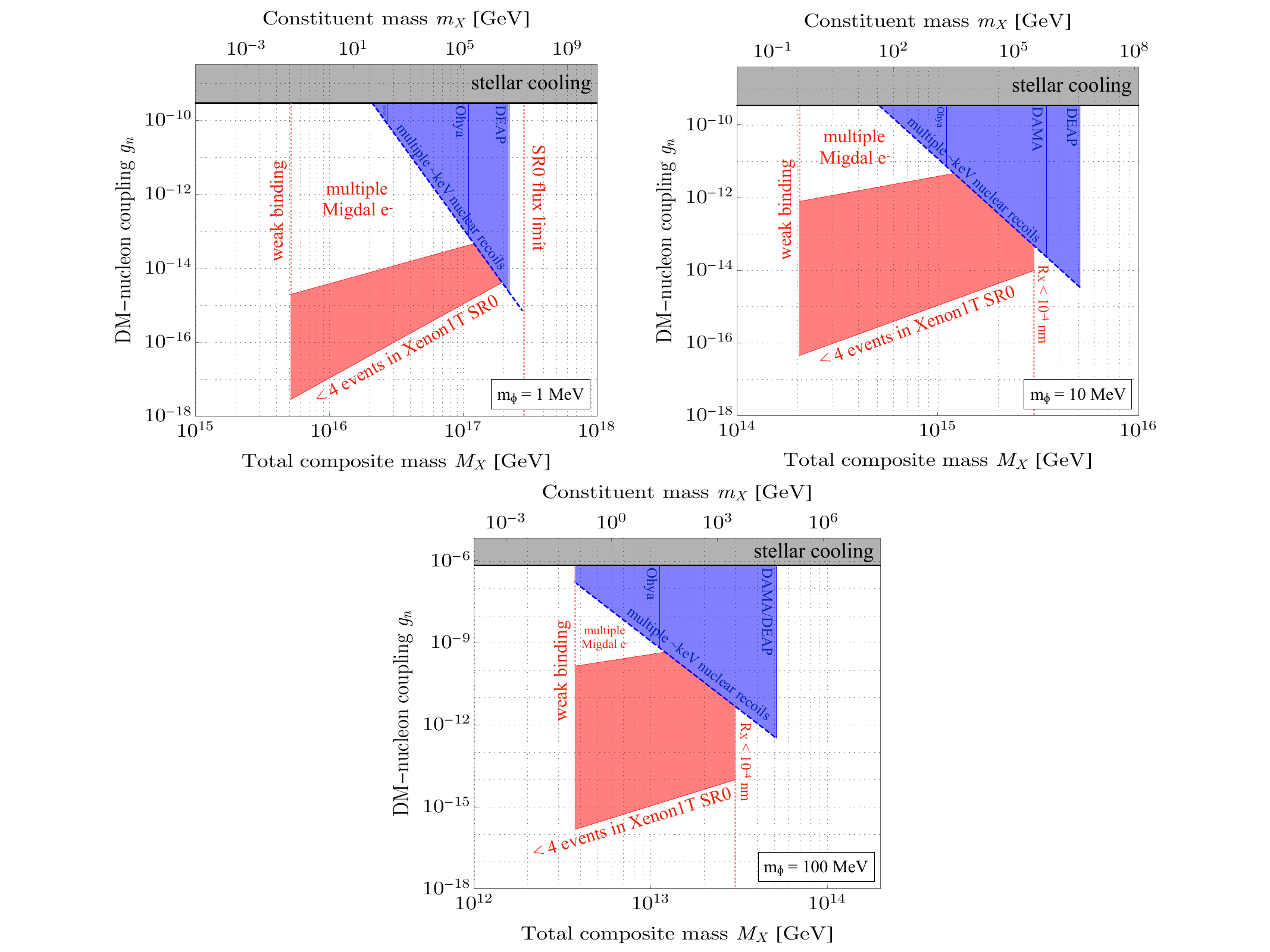}}
    
     \caption{Constraints on the DM-nucleon coupling from \acro{XENON-1T}'s first dark matter search (SR0) \cite{Aprile:2017iyp} as a function of composite mass (\textbf{red}), for a fixed dark coupling $\alpha_X = 0.3$, and binding mediator ($m_\phi$) masses as specified. The upper horizontal scale indicates the corresponding constituent mass. Note that these results assume no dilution during cosmological synthesis ($i.e.$ $\zeta = 1$ in Eq.~\eqref{eq:cosmo1}). Constraints are drawn from the expected number of Migdal electrons from composites in $98 \ \rm kg \ yr$ of exposure, Eq.~\eqref{eq:Migdaliontot}, compared to the few electronic recoil events observed. Bounds terminate at high $M_X$ due to the composites being so strongly bound that $R_X < 10^{-4} \ \rm nm$; we conservatively require $R_X > 10^{-4} \ \rm nm$ so that each transiting composite intercepts a few Xenon nuclei. At small $M_X$, the strong binding condition is not met. The line `SR0 flux limit' indicates the maximum mass reach of \acro{XENON-1T}'s first dark matter search. Above the upper line, a single composite will produce multiple Migdal electrons, requiring a new analysis of \acro{XENON-1T} data. The dashed blue line shows where each composite produces multiple $\sim \kev$ nuclear recoils, constrained by \acro{DAMA} \cite{Bernabei:1999ui}, analysis of \acro{DEAP-3600} data \cite{Adhikari:2021fum} and Ohya quarry / Skylab \cite{Bhoonah:2020fys} (\textbf{blue}). We also show stellar cooling bounds on light scalar fields coupled to nucleons \cite{Hardy:2016kme,Knapen:2017xzo} (\textbf{gray}).}
    \label{fig:migdalbounds}
    
\end{figure}

Figure~\ref{fig:migdalbounds} shows the constraints on the dark matter-nucleon coupling $g_n$ based on \acro{XENON-1T}'s first dark matter search results, for composites with different mediator masses that were synthesized in the early universe without a dilution stage, $cf.$ Eq.~\eqref{eq:cosmo1} with $\zeta = 1$. A composite synthesis including a dilution stage $\zeta \ll 1$ results in much heavier composites, limiting their detection prospects at traditional dark matter experiments due to the flux limit set by their running time and size (however, large neutrino observatories can still play a role in their detection, see \cite{Acevedo:2020avd}). The bounds in Fig.~\ref{fig:migdalbounds} are derived from the observation of $\mathcal{O}(1)$ electron recoil events during \acro{XENON-1T}'s first run. The composite mass range is limited either by the strong binding condition, $\bar{m}_X \ll m_X$, or the minimum composite radius considered here. The upper bound of the constraints is set by Eq.~\eqref{eq:Migdalionsingle}. For larger couplings, a single composite would create an ionization track while transiting the detection volume. While clearly detectable, such a signal requires a dedicated analysis of \acro{XENON-1T}'s data, which is out of the scope of this work. 

For comparison, we have also indicated the parameter space region where observable $\sim \kev$ nuclear recoils proceed. Within this region, we display existing constraints from \acro{DAMA} \cite{Bernabei:1999ui}, a recent analysis of multiscatter signatures of superheavy dark matter at the \acro{DEAP-3600} experiment \cite{Adhikari:2021fum}, and the Ohya quarry \cite{Orito:1990ny,Bhoonah:2020fys}, which are obtained by identifying the geometric cross section of the composite with a cross section for scattering with nuclei. Specifically, we use Eq.~\eqref{eq:sat-comp-radius} with Eq.~\eqref{eq:cosmo1} and the definition of $\bar m_X$ in the saturated regime to obtain the geometric cross-section for the composites, which is directly constrained by geometric composite nuclear scattering as detailed in \cite{Bhoonah:2020fys}. We note that there is a nontrivial relationship between the radius of the assembled composite and $m_X$; the scaling follows $R_X \propto m_X^{-3/5}$, meaning more massive composites with more massive constituents will be smaller in size. For this work, we have terminated the bounds at a cutoff radius $R_X \simeq 10^{-4} \ \rm nm$, to ensure that each composite encounters at least one xenon nucleus during its transit through \acro{XENON-1T}'s detection volume. Smaller composite geometric cross-sections can in principle be considered, but we have left a detailed treatment of this regime for future investigation. Additional bounds using nuclear scattering should be obtainable with new multiscatter analyses at experiments like \acro{XENON-1T} \cite{Bramante:2018qbc,Bramante:2018tos,Bramante:2019yss,Clark:2020mna}, as were recently obtained by \acro{DEAP-3600} \cite{Adhikari:2021fum}.  We also show existing stellar cooling constraints on the coupling \cite{Hardy:2016kme,Knapen:2017xzo}. Overall, these plots demonstrate that couplings well below existing experimental or stellar cooling constraints can be probed for large composite states by accounting for the Migdal effect. We also remark that for the regions we have constrained, the bounds are equally valid for a repulsive interaction between nuclei and the composite state, since the Migdal effect is independent of the direction of the momentum change. 

Finally, we comment on other existing constraints on nucleon couplings to light scalars from cosmological and experimental origin. As discussed extensively in \cite{Knapen:2017xzo}, the light scalar $\phi$ may contribute to the number of relativistic degrees of freedom at Big Bang Nucleosynthesis. However, for couplings $g_n \lesssim 10^{-9}$, its resulting contribution is not in tension with limits on $\Delta N_{\rm eff}^{\rm BBN}$ measured from primordial H and D abundances, as well as similar limits from CMB data and large-scale structure. In addition, one might consider whether the composite states themselves alter cosmological predictions like BBN; however cosmological observables appear to be unaltered by these composites, primarily because of their low number density \cite{Acevedo:2020avd}. There are also constraints on $g_n$ from accelerators. At present, rare $B$ and $K$ meson decays place limits of order $g_n \lesssim 10^{-8}$ \cite{Knapen:2017xzo}, which are less restrictive than the stellar cooling constraints we account for. 

\subsection{White Dwarf Explosions and Type-Ia Supernovae}
\label{sec:IaSNe}

Recent studies \cite{Scalzo:2014wxa,NearbySupernovaFactory:2014mcg} indicate that a sizable number of Type-Ia supernovae explode from sub-Chandrasekhar mass white dwarfs. However, the actual mechanism by which they proceed from such progenitors is not completely understood. Among the possible ignition scenarios, one can find matter accretion from a neighbouring star \cite{1982ApJ...253..798N,Han:2003uj,Wang:2018pac}, binary mergers \cite{doi:10.1111/j.1365-2966.2011.19361.x,doi:10.1093/mnras/stw1575} or helium shell ignition \cite{2041-8205-770-1-L8}. All of these processes require binary companions, whereas astronomical evidence indicates that a significant fraction of these events occur in single white dwarf systems \cite{2015Natur.521..332O,Maoz:2013hna}. In this context, there have been several proposals for dark sectors capable of igniting a thermonuclear runaway in single sub-Chansdrasekhar white dwarfs as possible solutions to this problem. These include accumulation and collapse of asymmetric dark matter cores that transfer their gravitational energy via dark matter-nucleus scattering \cite{Bramante:2015cua,Acevedo:2019gre,Janish:2019nkk}, evaporation to Hawking radiation of black holes formed from the collapsed dark matter cores \cite{Acevedo:2019gre,Janish:2019nkk}, heavy dark matter annihilation or decay to Standard Model particles \cite{Graham:2018efk}, pycnonuclear reactions enhanced by charged massive particles \cite{Fedderke:2019jur}, and the transit of primordial black holes \cite{Graham:2015apa,Montero-Camacho:2019jte} (see also \cite{Bertone:2007ae,McCullough:2010ai,Hooper:2010es,Amaro-Seoane:2015uny,Bramante:2017xlb,Dasgupta:2019juq,Curtin:2020tkm,Horowitz:2020axx,Chan:2020ijt,Dasgupta:2020mqg,Bell:2021fye} for related work on dark matter in white dwarf stars). 

Recently in Ref.~\cite{Acevedo:2020avd} we identified how large DM composites could spark the ignition of white dwarfs explosions, and indicated relevant DM composite parameter space. In this Section we provide a more detailed exposition, and examine how saturated composites produce enough localized heating via exothermic fusion reactions occurring in the DM composite interior, to initiate a thermonuclear runaway at the center of a carbon-oxygen (C/O) white dwarf. For white dwarf ignition conditions, we follow the numerical results from \cite{1992Timmes}, where different critical temperatures and trigger masses were numerically computed for a set of compositions. A critical temperature $T_{c} \sim 10^{10} \ {\rm K} \sim 1 \ \mev$ is typically sufficient to satisfy ignition for any composition and trigger mass. Therefore, for ignition the composite potential must be $\langle \varphi \rangle=Ag_{n}(m_{X}/g_{X}) \gtrsim 1 \ \mev$, as this quantity determines the kinetic energy of nuclei inside the composite. The carbon and oxygen nuclei heated to the critical temperature inside the composite state will fuse in (mostly) exothermic reactions. For a runaway fusion reaction to occur, the heat released must overcome the rate at which it is dissipated by the white dwarf material. As we will show below, this condition imposes a minimum composite size in order to have a sufficiently large number of reactions proceeding in the composite interior.

\sloppy We first comment on the kinetic state of nuclei in the white dwarf, in relation to the transiting composites. Neglecting the initial halo velocity, dark matter composites will cross the white dwarf at approximately escape velocity, \mbox{$v_{\rm esc}=(2GM_{*}/R_{*})^{1/2} \simeq 3.5 \times 10^{-2} \ (M_*/1.3M_\odot)^{1/2} \ (R_* / 3000 \ \rm km)^{-1/2}$}, where $M_\odot$ is a solar mass. We can directly compare this speed to the motion of the stellar constituents. For cold $\sim$Gyr old white dwarfs, nuclei are arranged in a crystalline lattice, and we assume they vibrate in their zero point state $|0\rangle$ with a natural frequency given by the plasma frequency of the medium, 

     \begin{equation}
       \Omega_{p}=\left(\frac{4 \pi Z^{2} e^{2} n_{N}}{m_{N}}\right)^\frac{1}{2} \simeq 37 \ \kev \left(\frac{\rho_*}{10^{9} \ \rm g \ cm^{-3} }\right)^\frac{1}{2} \left(\frac{Z}{6}\right)\left(\frac{A}{12}\right)^{-1}.
       \label{eq:plasmafreq}
      \end{equation}

In such a state, their average velocity is $\langle0| v |0\rangle = 0$, whereas the velocity dispersion is \mbox{$\left(\langle0| v^{2} |0\rangle\right)^{1/2} = \left(\Omega_{p}/m_{N}\right)^{1/2} \simeq 10^{-3} \ll v_{\rm esc}$}, for $m_N \simeq 12$ - $16 \ \gev$. Therefore, we safely assume nuclei to be at rest compared to the DM composite velocity in the stellar rest frame.

\subsubsection{Composite energy loss during white dwarf transit}

Between the DM composite and the white dwarf medium, there are a number of energy transfer rates to consider: 1) the nuclear energy release rate from thermonuclear reactions from nuclei fusing inside the composite, and 2) the energy loss rate of accelerated nuclei from the DM composite constituents scattering with nuclei. As will see in Section \ref{sec:nucdmint}, the interactions between nuclei and dark matter constituents inside the composite are severely suppressed. However, the nuclei accelerated inside the composite can instead scatter against the degenerate electrons in the white dwarf medium, which efficiently conduct heat. In addition, these transiting nuclei can also emit photons and neutrinos.

Since nuclei losing kinetic energy results in the composite slowing down,  as discussed in Appendix~\ref{sec:app-a}, these processes can potentially halt the composite transit through the star, preventing ignition at the center. Now we discuss these energy loss channels in detail:

\begin{itemize}
    \item \textit{Electron conduction}: Accelerated nuclei inside the composite will scatter against degenerate electrons in the white dwarf medium. Since electrons are relativistic for stellar densities $\rho_* \gtrsim 10^6 \ \rm g \ cm^{-3}$, these are highly efficient heat carriers. This makes the stellar material beneath the radiative envelope effectively isothermal, with temperatures $T_{*}\sim (10^{6}$ - $10^{7}) \ \rm K$ for $\sim$Gyr old white dwarfs. The heat conduction rate out of a spherical region of stellar material of radius $R_X$ at the critical temperature $T_c$ is given by \cite{Potekhin:1999yv},
    
    \begin{align}
        \dot{Q}_{\rm cond}= & \frac{4 \pi^{2} R_X T_c^{3}(T_c-T_{*})}{15 \kappa_{c} \rho_*} \simeq \frac{4 \pi^{2} T_c^{4} R_{X}}{15 \kappa_{c} \rho_*} \nonumber \\ & \simeq{{10^{27}} \ \gev \ {\rm s^{-1}} \ \left(\frac{\rho_{*}}{10^{9} \ \rm g \ cm^{-3}}\right)^{\frac{4}{15}}} \left(\frac{R_{X}}{\rm{ \mu m}}\right),
     \label{eq:wdQdiff}
    \end{align}
    
    where $\kappa_{c}\simeq 10^{-9} \ {\rm cm^{2} \ g^{-1}} \ (T/10^{7} \ {\rm K})^{2.8} \ (\rho_*/10^{9} \ \rm g \ cm^{-3})^{-1.6}$ is the conductive opacity of the relativistic white dwarf electrons \cite{Potekhin:1999yv}. In the second equality, we have neglected the stellar background temperature compared to the critical ignition temperature. 
    
    \item \textit{Photon emission}: The heated stellar material inside the composite will also cool down by radiating photons. Due to the high opacity of the stellar material, photons will scatter many times inside the composite, thermalizing with the heated plasma. Therefore, we assume this cooling process can be modeled using a blackbody radiation spectrum at temperature $T_c$. The rate at which energy is radiated out of a composite of radius $R_X$ is \cite{1983bhwd.book.....S},
    
    \begin{align}
        \dot{Q}_{\rm rad} = & \frac{4\pi R_X^2 \sigma_{\rm SB} \nabla T^4}{\kappa_r \rho_*} \simeq \frac{16\pi R_X^2\sigma_{\rm SB} m_\phi T_c^4}{\kappa_r \rho_*} \nonumber \\ & \simeq 10^{24} \ {\rm GeV \ s^{-1}} \left(\frac{m_\phi}{\rm keV}\right) \left(\frac{R_X}{\rm \mu m}\right)^2,
        \label{eq:wdQrad}
    \end{align}
    
    where $\sigma_{\rm SB} = \pi^2 / 60$ is the Stefan-Boltzmann constant in natural units, and $\kappa_r \simeq 10^{7} \ {\rm cm^2 \ g^{-1}} \ (T_*/10^7 \ \rm K)^{-7/2} \ (\rho_*/10^9 \ {\rm g \ cm^{-3}})$ is the white dwarf radiative opacity for free-free electron transitions \cite{Kippenhahn:1994wva,Fedderke:2019jur}. In the second equality, we have approximated the temperature gradient $\nabla T^4 = 4 T^3 \nabla T \simeq 4 T_c^3 \left((T_c-T_*)/m^{-1}_\phi\right) \simeq 4 m_\phi T_c^4$. This approximation is accurate for saturated composites because $m_{\phi}^{-1} \ll R_X$, so the potential is short-ranged compared to the composite size.
    
    \item \textit{Neutrino emission}: At the critical temperature $T_c$, neutrino emission is dominated by electron-positron annihilation, with an emission rate of $10^{29}$ - $10^{30} \ \rm GeV \ cm^{-3} \ s^{-1}$ in the stellar density range $(10^8$ - $10^{10}) \ \rm g \ cm^{-3}$ \cite{1987ApJ...313..531S,1989ApJ...339..354I,1996ApJS..102..411I}. The white dwarf material is transparent to neutrinos of $\sim \mev$ energies, and so they escape the star carrying all of their energy away. This yields a neutrino energy loss rate of order $\dot{Q}_{\nu} \simeq 10^{18} \ {\rm GeV \ s^{-1}} \ (R_X/{\rm \mu m})^3$.
\end{itemize}

Comparing these energy loss mechanisms, we see that photon and neutrino losses will be sub-dominant compared to electron conduction unless composites have radii $R_X \gtrsim 0.1 \ \rm cm$ and $R_X \gtrsim 100 \ \rm cm$ respectively. However as we detail below, composites in excess of $R_X \gtrsim 1 \ \rm cm$ will be too scarce in DM halos and therefore unable to explode white dwarfs on $\sim$Gyr timescales. Thus, we are only concerned with the energy loss from conduction and, at the highest composite masses, radiation. 

We now turn to the question of whether composites will be stopped through dissipative processes as they cross the white dwarf. Let us compare the above energy loss rates to the composite kinetic energy at the surface of the white dwarf, and the white dwarf crossing time. The initial composite kinetic energy at the star surface is

\begin{equation}
    \frac{1}{2} M_X v_{\rm esc}^2 \simeq 10^{27} \ \gev \left(\frac{M_X}{10^{30} \ \gev}\right) \left(\frac{v_{\rm esc}}{3 \times 10^{-2}}\right)^2
    \label{eq:wdKi}
\end{equation}
For sub-Chandrasekhar white dwarfs, with radii ranging $(2500$ - $3000) \ \rm km$, composites will cross the star in $\sim 1 \ \rm s$. Thus, comparing Eq.~\eqref{eq:wdKi} to the conduction and radiation losses above, we see that composites of masses $M_X \gtrsim 10^{30} \ \gev$ will not be significantly stopped by these dissipation processes. 

\subsubsection{Composite heating of white dwarf material}

Next, to determine whether the composite can cause a thermonuclear runaway in the white dwarf, we must compare the above composite cooling rates to a number of timescales relevant for thermonuclear reactions inside the composite as it crosses the white dwarf. First we look at the timescale for stellar material to be heated as it crosses the DM composite's boundary. The heat capacity of the white dwarf ions is $C_{v}\simeq 2\pi \rho_* R_{X}^{3}/m_N$. Using the acceleration timescale Eq.~\eqref{eq:accel1}, we estimate the heating rate of the stellar material from accelerating across the composite boundary to be

\begin{equation}
 \dot{Q}_{\rm nuc}\simeq C_{v} \ T_c \ \tau_{\rm accel}^{-1} \simeq 10^{34} \ \gev \ {\rm s^{-1}}  \left(\frac{v_{X}}{10^{-2}}\right) \left(\frac{m_\phi}{\kev}\right) \left(\frac{\rho_{*}}{10^{9} \ \rm g \ cm^{-3}}\right) \left(\frac{R_{X}}{\rm{\mu m}}\right)^{2}.
 \label{eq:wdQnuc1}
 \end{equation} 
This is much larger than the above cooling rates for the parameters indicated. Hence, for composites with radii $R_X \gtrsim 10^{-2} \ \rm \mu m$, $\dot{Q}_{\rm nuc} \gg \dot{Q}_{\rm cond},\dot{Q}_{\rm rad}$, and so neither conductive nor radiative losses prevent stellar material from being rapidly heated to the critical ignition temperature during the composite transit.

The same cooling mechanisms discussed above must also be compared to the timescale required for the thermonuclear runaway itself. Specifically, to ignite the white dwarf star we must require the heat diffusion rate to fall below the rate at which energy is released from the thermonuclear reactions. At the critical temperature $T_{c}$, the ions can be treated as an ideal gas and nuclear reactions proceed in the classical thermonuclear regime, $i.e.$ screening effects are negligible. For a C/O white dwarf, there will be three primary fusion processes we will consider: $\rm ^{12}C+^{12}\!C$, $\rm ^{12}C+^{16}\!O$ and $\rm ^{16}O+^{16}\!O$. These reactions respectively produce the compound nuclei $\rm ^{24}Mg$, $\rm ^{28}Si$ and $\rm ^{32}S$, each decaying via n-, p- and $\alpha$- channels. The nuclear burning rate is set by the astrophysical S-factor which determines the fusion cross section at low energies, as well as the Coulomb barrier penetration factor, which are extensively discussed in \cite{Gasques:2005ar,Yakovlev:2006fi}. Using the formalism outlined in these References, we find the specific rates for each reaction per unit volume,

\begin{equation}
    \frac{d\dot{R}_{\rm C+C}}{dV}\bigg|_{T_c} \simeq 10^{43} \ \rm cm^{-3} \ s^{-1} \left(\frac{X_C}{0.5}\right)^2 \left(\frac{\rho_*}{10^9 \ \rm g \ cm^{-3}}\right)^2,
\end{equation}

\begin{equation}
    \frac{d\dot{R}_{\rm C+O}}{dV}\bigg|_{T_c} \simeq 10^{42} \ \rm cm^{-3} \ s^{-1} \left(\frac{X_C}{0.5}\right) \left(\frac{X_O}{0.5}\right) \left(\frac{\rho_*}{10^9 \ \rm g \ cm^{-3}}\right)^2,
\end{equation}

\begin{equation}
    \frac{d\dot{R}_{\rm O+O}}{dV}\bigg|_{T_c} \simeq 10^{40} \ \rm cm^{-3} \ s^{-1} \left(\frac{X_O}{0.5}\right)^2 \left(\frac{\rho_*}{10^9 \ \rm g \ cm^{-3}}\right)^2,
\end{equation}
where $X_C + X_O = 1$ are the mass fractions of carbon and oxygen. Each fusion process releases $Q_{\rm C+C} \simeq 3.16 \ \mev$, $Q_{\rm C+O} \simeq 6.51 \ \mev$ and $Q_{\rm O+O} \simeq 13.09 \ \mev$ of average heat per reaction \cite{Caughlan:1987qf}. The subsequent nuclear energy release rate is,

\begin{equation}
    \dot{Q}_{\rm fus}\simeq \frac{4 \pi}{3} R_X^3 \left(\ \sum_{j= \rm CC,CO,OO} Q_{j} \frac{d\dot{R}_{j}}{dV}\bigg|_{T_{c}}\ \right)
    \label{eq:wdQfus}
\end{equation}

The nuclear energy release rate Eq.~\eqref{eq:wdQfus} scales as $R_{X}^{3}$, whereas the heat conduction rate Eq.~\eqref{eq:wdQdiff} only scales as $R_{X}$. This implies that there is a minimum composite radius for which the nuclear energy released exceeds the conduction rate and the runaway commences. For concreteness, we consider two benchmark cases: a pure $X_{\rm C} = 1$ composition, and an $X_{\rm C} = X_{\rm O} = 0.5$ C/O mixture. Using the reaction rates above, we obtain a nuclear energy release rate of $\dot{Q}_{\rm fus} \simeq 10^{33} \ {\rm GeV \ s^{-1}} \ (\rho_*/10^9 \ {\rm g \ cm^{-3}})^2 (R_X/\rm \mu m)^3$ for pure carbon, and $\dot{Q}_{\rm fus} \simeq 4 \times 10^{32} \ {\rm GeV \ s^{-1}} \ (\rho_*/10^9 \ {\rm g \ cm^{-3}})^2 (R_X/\rm \mu m)^3$ for a C/O mixture. For both benchmark compositions, composites with radii above $R_{X} \gtrsim 10^{-2} \ \rm \mu m$ will satisfy $\dot{Q}_{\rm nuc} \gg \dot{Q}_{\rm cond},\dot{Q}_{\rm rad}$. Using Eq.~\eqref{eq:sat-comp-radius}, this implies a minimum composite mass $M_X \gtrsim 10^{23} \ \gev$, which is smaller than the required mass for the composites to reach the stellar core, $cf.$ Eq.~\eqref{eq:wdKi} and surrounding discussion. Thus, we conclude that composites with masses and radii

\begin{equation}
    M_X \gtrsim 10^{30} \ \gev \ \ \ , \ \ \ R_X \gtrsim 10^{-2} \ \rm \mu m
\end{equation}
are capable of reaching the core of a massive C/O white dwarf and ignite a nuclear runaway, so long as their coupling to SM nucleons is sufficiently strong. Such minimum coupling is determined from the critical temperature by setting Eq.~\eqref{eq:accelsr3} equal to $T_c$. Following this procedure, the minimum coupling for igniting a thermonuclear runaway in the white dwarf is

\begin{equation}
    g_n \gtrsim 10^{-12} \left(\frac{m_X}{10^8 \ \gev}\right)^{-1}.
    \label{eq:wdbounds}
\end{equation}
We remark that given stellar cooling constraints, which already bound the coupling $g_n \lesssim 10^{-12}$ for mediators with masses $\lesssim \kev$, typically it will be composite states made of heavy $m_X \gg \rm PeV$ constituents, which have a sufficiently strong potential, that will cause a white dwarf star to explode, as we explored in \cite{Acevedo:2020avd}. Thus, a cosmological synthesis with a dilution stage is necessary to synthesize composites with an adequate size and coupling to produce Type-Ia supernovae, $cf.$ $\zeta \ll 1$ in Eq.~\eqref{eq:cosmo1}.

The existence of old white dwarfs in the mass range $(1.1$ - $1.4) \ M_\odot$, corresponding to central densities $\rho_* \simeq (10^{8}$ - $10^{10}) \ \rm g \ cm^{-3}$, implies constraints on the nucleon-dark matter coupling $g_n$, assuming at least one encounter with a composite occurred during their $\sim$Gyr lifespan. Although some fraction of these white dwarfs will possess O/Ne cores, which have not been discussed here, we remark that multiple studies \cite{2017A&A...602A..16T,Maoz:2018epf,2020A&A...636A..31T} indicate that an $\mathcal{O}(1)$ fraction of massive C/O white dwarfs are formed as the result of binary mergers \cite{Yoon:2007pw,LorenAguilar:2009cv}. Furthermore, there exist single-evolution channels by which massive C/O white dwarfs may form, such as reduced mass loss rates in the asymptotic giant branch phase \cite{2019NatAs...3..408D,2021A&A...646A..30A} or enhanced rotation \cite{1996ApJ...472..783D,2021A&A...646A..30A}. Searching the Montreal White Dwarf Database (MWDD) \cite{2017ASPC}, we find an $\sim \mathcal{O}(10^3)$ sample of single white dwarfs in the mass range of interest, with cooling ages ranging $1$ - $5$ Gyr at distances $\lesssim 10^3 \ \rm pc$. Based on the above considerations, a fraction of this sample must be C/O white dwarfs that have not exploded, implying that either ignition conditions were not satisfied, or else they have not encountered a composite in their lifetime. The former condition constrains $g_n$ according to Eq.~\eqref{eq:wdbounds}, whereas the latter condition implies that such constraints are valid for composites with masses $M_X \lesssim 10^{42} \ \gev$, assuming a dark matter halo density $\rho_{X} \simeq 0.4 \ \rm GeV \ cm^{-3}$. We emphasize that any C/O white dwarf within this mass range will impose a similar condition on the minimum mass and radius of ignition-capable composites, since the central density spans $\sim 2$ orders of magnitude, resulting in an $\sim \mathcal{O}(1)$ correction factor to the minimum composite radius for ignition. 

\subsection{Terrestrial Heat Flow}
\label{sec:earthflow}

A number of works have studied the impact of dark matter on planets and their satellites \cite{Kawasaki:1991eu,Mitra:2004fh,Mack:2007xj,Adler:2008ky,Chauhan:2016joa,Bramante:2019fhi,Garani:2019rcb,Chan:2020vsr,Leane:2020wob,Acevedo:2020gro,Leane:2021tjj}. In particular, bounds have been placed on different dark matter models based on the heat that captured dark matter would produce in the Earth's core from annihilation \cite{Mack:2007xj,Chauhan:2016joa,Bramante:2019fhi} or gravitational collapse into evaporating black holes \cite{Acevedo:2020gro}, which would exceed the $\dot{Q}_\oplus \lesssim 44 \ \rm TW \simeq 10^{23} \ \rm GeV \ s^{-1}$ of heat flow from the surface of the Earth \cite{williams1974,lister1990}. In this Section, we discuss the heat signature produced by large composites in the Earth's mantle and core, and argue that although a significant fraction of the composite dark matter flux can be gravitationally captured by the Earth, these composites do not produce a heat flow comparable to the value currently measured.

Composite dark matter can be captured by the Earth's crust and mantle from the various energy dissipation mechanisms which, as detailed in Appendix~\ref{sec:app-a}, translates into lost kinetic energy as it travels through matter.  The composite dark matter number flux through the Earth is

\begin{equation}
    \Phi_X = 2\pi R_\oplus^2 \langle v_X \rangle \frac{\rho_X}{M_X} \simeq 10^{5}\textrm{ s}^{-1} \left( \frac{M_X}{10^{20} \ \gev} \right)^{-1},
    \label{eq:eanumFlux}
\end{equation}
where $R_\oplus = 6371 \ \rm km$ is the radius of the Earth, $\langle v_X \rangle\simeq 300 \rm \ km \ s^{-1}$ is the average DM velocity entering the Earth, and $\rho_X \simeq 0.4$ GeV/cm$^3$ is the local DM mass density \cite{Mack:2007xj,Iocco:2011jz,Buch:2018qdr,Bramante:2019fhi}. By itself, the kinetic energy of this DM flux falls well short of the observed heat flow of the Earth. Each composite has a kinetic energy of order $\sim 10^{13} \ \gev \ (M_X / 10^{20} \ \gev)(v_X/ 300 \ \rm km \ s^{-1})^2$, in contrast with the $\sim \dot{Q}_\oplus / {\Phi_X} \simeq 10^{18} \ \gev \ (M_X / 10^{20} \ \gev)$ required per composite to produce an observable deviation in the heat flow from Earth. However, we must consider if the composites that are captured release further energy into the core and mantle through nuclear collisions and reactions in their interiors. At temperatures $\gtrsim 100 \ \rm keV$, thermal Bremsstrahlung will significantly stop composites in the mantle \cite{Acevedo:2020avd}, which will then slowly drift and settle at the core, on a timescale that can be computed using the methods described in \cite{Bramante:2019fhi,Acevedo:2020gro}. Given that the mantle composition is predominantly $^{16}$O, we expect sizeable oxygen burning reactions to occur at such temperatures. Once the composites have settled at the core, they could later release heat as matter is accumulated in their interior, since matter they accumulate will have fallen into the composite's potential, and would radiate a corresponding amount of potential energy. Finally, we should also consider whether these dark matter states might congregate at the center of the Earth and gravitationally collapse into a black hole light enough to evaporate via Hawking radiation, provided enough dark matter is captured \cite{Acevedo:2020gro}. Below we detail each of these processes:

\begin{itemize}

    \item \textit{Fusion reactions:} The captured composites could release heat via fusion reactions occurring in their interiors. The Earth's mantle is mostly composed of $^{16}$O and silicon, while the core is mostly composed of $^{56}$Fe, which can no longer fuse and therefore is not considered in this analysis. Oxygen burning is the only plausible reaction that can occur at temperatures $T \sim 100 \ \kev$ - $1 \ \mev$, with a highly temperature-dependent rate we extract from \cite{Caughlan:1987qf}. To achieve such temperatures, given that stellar cooling constraints limit the dark matter-nucleon coupling to $g_n \lesssim 10^{-12}$ - $10^{-10}$ in the eV - GeV mediator range, constituent masses must be in excess of $m_X \gtrsim 10^6 \ \gev$. In $t_\oplus \simeq 4.5 \ {\rm Gyr} \simeq 10^{17} \ {\rm s}$, about $10^{22} \ (M_X / 10^{20} \ \gev)^{-1}$ composites are captured. Each of them must then release at least $10 \ {\rm GeV \ s^{-1}} \ (M_X / 10^{20} \ \gev)$ to produce $\dot{Q}_{\oplus}$. However, the energy released from oxygen burning reactions per composite is $\sim 10^{-1} \ {\rm GeV \ s^{-1}} \ (M_X / 10^{20} \ \gev) (\bar{m}_X/5 \ \gev)^{-4}$. Thus, oxygen fusion reactions in the mantle triggered by these composites cannot substantially alter the observed Earth heat flow.

    \item \textit{Gravitational collapse}: The composites accumulated in the Earth's core will eventually fuse into a single larger composite state. We do not discuss the timescale for such process here. Instead, we point out that if the critical mass is exceeded, the dark matter will collapse into a black hole which can either overheat or destroy the Earth depending on its initial mass, as discussed extensively in \cite{Acevedo:2020gro}. However, the critical mass for collapse is $M_{\rm coll} \simeq M_{pl}^3 / \bar{m}_X^2 \simeq 10^{55} \ \gev \ (\bar{m}_X / 5 \ \gev)^{-2}$ \cite{Gresham:2018rqo}. The total dark matter mass flux, on the other hand, is $2\pi R_\oplus^2 \langle v_X \rangle \rho_X \simeq 10^{25} \ \rm GeV \ s^{-1}$, $cf.$ Eq.~\eqref{eq:eanumFlux}, and therefore insufficient for the critical mass to be accumulated in $t_\oplus \simeq 4.5 \ {\rm Gyr} \simeq 10^{17} \ {\rm s}$.
    
        \item \textit{Matter compression:} A detailed study of the phase of matter after a captured composite has settled is out of the scope of this work. A simple estimate, however, indicates that the heat released as nuclei settle in the composite interior is insufficient to produce $\dot{Q}_{\oplus}$. As specified above, each captured composite would need to deposit $\sim \dot{Q}_\oplus / {\Phi_X} \simeq 10^{18} \ \gev \ (M_X / 10^{20} \ \gev)$ in the form of compressional heating once they reach the core and thermalize. By contrast, nuclei would ultimately release an energy of order $\langle \varphi \rangle$ in the form of heat, resulting in a total energy deposition of $n_N \langle \varphi \rangle R_X^3 \sim \mev \ (n_N / 10^{23} \ {\rm cm^{-3}})(\langle \varphi \rangle/\mev)(M_X / 10^{20} \ \gev) (\bar{m}_X/5 \ \gev)^{-4}$, where the number density has been normalized to the Earth's inner core value. However, it is probably the case that Standard Model nuclear material collected into the composite will continue to accumulate, potentially reaching a density well beyond $10^{23} \ {\rm cm^{-3}}$, until it is stabilized against further accumulation by electron or nucleon degeneracy pressure. Put differently, we should consider whether nuclei continue to collect until the interior of DM composites resembles ``white dwarf'' or ``neutron star'' material. In this case, the above estimates indicate that composites with an extremely small in-medium mass ($e.g.$ $\bar{m}_X \ll 5 ~\mev$, assuming a $10^{32} ~ {\rm cm^{-3}}$ ``WD'' nuclear density) may appreciably heat the Earth. We leave a proper study of this phenomenon to future work.

\end{itemize}

From these estimates we conclude that while dark matter composites captured by the Earth may be responsible for a fraction of the heat output, they cannot account for the total flow observed, via fusion reaction of gravitational collapse. On the other hand, we found that for ``matter compression'' a bound might be attained for larger composites than those we consider, in the case that the accumulated nuclear material inside the composite comes to exceed terrestrial densities by many order of magnitude. We have left this possibility open to future inquiry, since this will depend on the eventual equilibrium SM nuclear state of very large DM composites inside the Earth. Before concluding, we proceed to examine interactions between SM nuclei and constituents inside DM composites.

\section{Scattering Interactions of Nuclei with Constituents}
\label{sec:nucdmint}

In this Section, we discuss a few different interactions between nuclei and dark matter constituents, $i.e.$ processes that fall under the fourth category depicted in Fig.~\ref{fig:DMschem}. These processes are of some interest, since if nuclei lose kinetic energy while scattering in the composite, this will cause a kinetic drag on the composite state, resulting in a stopping force as the composite moves through matter (for more discussion see Appendix~\ref{sec:app-a}). However, we will find that the effect of constituent scattering processes is extremely suppressed for degenerate dark matter fermionic composites, relative to the nuclear acceleration processes detailed above. Nevertheless, for completeness we hereafter analyze coherent composite-nucleus scattering as well as up-scattering of individual dark matter constituents above the Fermi level. In addition, this calculations may prove more relevant for less degenerate composites not considered here. In Appendix~\ref{sec:app-b}, we also comment on nuclear interactions with low-lying collective excitations, which we also find to be negligible for the composite parameter space considered in this study.

\subsection{Coherent Composite-Nucleus Scattering}
\label{sec:nucdmint1}

We first consider the scattering process $^{N_X} \! X(\mathbf{p})+ N(\mathbf{k}) \  \rightarrow \  ^{N_X} \! X(\mathbf{p'})+ N(\mathbf{k'})$ whereby a composite with $N_{X}$ constituents and radius $R_{X}$ coherently scatters against a nucleus and transfers a fraction of its total momentum and kinetic energy. The cross section for this process can be parametrized in terms of the individual scattering cross section against constituents, multiplied by adequate form factors that encapsulate the nuclear and composite substructure \cite{Hardy:2015boa,Coskuner:2018are},

\begin{equation}
\left(\frac{d^2\bar{\sigma}}{dq d\omega}\right)_{XN\rightarrow XN}= \bar{\sigma}_{0} \left(\frac{q}{2 v_{X}^{2} m_{N}^{2}}\right) S(q,\omega)
\label{eq:cstot1}
\end{equation}
where $S(q,\omega)=A^{2} \ |F_{a}(q r_{N})|^{2} \ \delta(\omega-q^{2}/2m_{N})$ is the nuclear structure factor. This function sets the dispersion relation between momentum and energy exchanged in the non-relativistic limit. The variables $\omega$ and $q$ are respectively the energy and momentum transfer in the collision. Assuming the nucleus is initially at rest, the maximum momentum exchange is $q_{\rm max} \simeq 2 m_N v_X$ in the limit $m_N \ll M_X$. The function 

\begin{equation}
F_{a}(q r_{N})=\frac{3 j_{1}(qr_{N})}{qr_{N}} e^{-q^{2}r_{N}^{2}},
\end{equation}
is the nuclear Helm form factor \cite{Helm:1956zz,Lewin:1995rx} that accounts for nuclear substructure, where $j_{1}$ is a spherical bessel function of first kind and $r_{N} \simeq (1.25 \ {\rm fm}) A^{1/3}$ is the nuclear radius.

The constant $\bar{\sigma}_{0}$ in Eq.~\eqref{eq:cstot1} reads 

\begin{equation}
\bar{\sigma}_{0}=\frac{g_{n}^{2}g_{X}^{2} m_{N}^{2}}{4\pi \tilde{m}_{\phi}^{4}},
\label{eq:csref1}
\end{equation}
and is a reference cross-section for point-like dark matter scattering against a free nucleus in the limit $m_{N} \ll M_{X}$. This reference cross-section includes the mediator mass correction due to screening effects, $\tilde{m}_\phi^2 = m_\phi^2 + \delta m_\phi^2$ with $\delta m_\phi^2 \simeq \left(4 \pi \alpha_X \bar{m}_X^4 \right)^{1/2}$. Eq.~\eqref{eq:cstot1} therefore corresponds to a differential cross section for point-like dark matter with the mass and couplings of a composite. This cross section is related to the cross section for an extended composite state through multiplication of additional form factors associated with the composite internal structure \cite{Hardy:2015boa,Coskuner:2018are},

\begin{equation}
 \left(\frac{d^2\sigma}{dq d\omega}\right)_{XN\rightarrow XN}= N_{X}^{2} f^2(\Lambda) |F_{X}(qR_{X})|^{2} \times \left(\frac{d^2\bar{\sigma}}{dq d\omega}\right)_{XN\rightarrow XN},
 \label{eq:cstot2}
\end{equation} 
where the $N_{X}$ factor explicitly contains the coherent enhancement. The function $f(\Lambda)$ is, 

\begin{equation}
    f(\Lambda) = \min \left[1,\left(\frac{\Lambda}{R_X}\right)^3\right],
    \label{eq:blob}
\end{equation}
where ratio $(\Lambda/R_{X})^{3}$ accounts for the partial overlap of the nucleus wave function with the composite state \cite{Grabowska:2018lnd}, and $\Lambda$ is the spatial spread of the nucleus. In other words, when $\Lambda \ll R_{X}$, the reduced wave function extent of the scatterer compared to the size of the composite state effectively limits the number targets for coherent scattering. For the composite sizes we consider here, which are comparable to the atomic separation in any medium, this suppression is significant. The extra form factor above is

\begin{equation}
F_{X}(qR_{X})=\frac{3 j_{1}(qR_{X})}{qR_{X}},
\end{equation}

and accounts for the composite structure. Note that $F_{X}$ corresponds to a sphere of homogeneous density, multiplied by $N_{X}$ which explicitly contains the coherent enhancement. 

To estimate the energy loss from composite-nucleus scattering as the composite goes through a material medium, we proceed to integrate Eq.~\eqref{eq:cstot2} over the energy transfer $\omega$,

\begin{equation}
\left(\frac{d\sigma}{dq}\right)_{XN\rightarrow XN} = A^2 N_{X}^{2} f^2(\Lambda) \bar{\sigma}_{0} \left(\frac{q}{2 m_{N}^{2} v_{X}^{2}}\right) \ |F_{X}(qR_{X})|^{2} \ |F_{a}(q r_{N})|^{2}
\label{eq:cstot3}
\end{equation}
The energy loss per unit distance travelled by the composite in a given medium is obtained by integrating Eq.~\eqref{eq:cstot3} over all possible momentum exchanges up to $q_{\rm max} \simeq 2 m_N v_X$, and multiplying by the target density $n_N$,

\begin{equation}
\left(\frac{dE}{dx}\right)_{XN \rightarrow XN} = n_{N} \int_{0}^{q_{\rm max}} dq \ \frac{q^2}{2m_N} \ \left(\frac{d\sigma}{dq}\right)_{XN\rightarrow XN}.
\end{equation}
The integration of the form factors above is numerically performed for each composite radius and target mass. Finally, we discuss the choice of the spatial extent $\Lambda$ for the scatterer wavefunction, $cf.$ Eq.~\eqref{eq:cstot3}. We consider two separate cases in this work:

\begin{itemize}
    \item \textit{Cold dense plasma:} This is the case for white dwarfs that have undergone crystallization. Electrons are fully ionized and form a degenerate gas that provides a neutralizing background, while nuclei are arranged in a lattice and vibrate at their zero point states with a natural frequency determined by the plasma frequency, $cf.$ Eq.~\eqref{eq:plasmafreq}. Therefore, assuming nuclei occupy the ground state $|0\rangle$ of a harmonic oscillator potential, we expect their initial wave function to be gaussian with a spatial spread given by

     \begin{equation}
       \Lambda_{\rm wd} = \left(\langle0| r^{2} |0\rangle\right)^\frac{1}{2} = \left(\frac{1}{2 m_{N} \Omega_{p}}\right)^\frac{1}{2} \simeq 4 \times 10^{-12} \ {\rm{cm}} \ \left(\frac{6}{Z}\right)^{\frac{1}{2}} \left(\frac{10^{6} \ \rm g \ cm^{-3}}{\rho_{*}}\right)^{\frac{1}{4}}.
      \end{equation} 
      
    As we have shown in Section~\ref{sec:IaSNe}, in order to ignite a carbon-oxygen sub-Chandrasekhar white dwarf, composite radii must be $R_{X} \gtrsim 10^{-2} \ \rm \mu m$. Comparing this to the length scale above, we see that the ratio $(\Lambda/R_{X})^{3} \lesssim 10^{-21}$, so the coherent enhancement factor $N_X^2$ is substantially reduced.
      
    \item \textit{Earth matter:} For matter at low densities and pressures, we approximate the wavefunction spread $\Lambda$ by the thermal de Broglie wavelength at a temperature $T$,

     \begin{align}
      \Lambda_{\rm th}=\left(\frac{2 \pi}{m_{N} T}\right)^\frac{1}{2} \, & \simeq 10^{-9}  \ {\rm{cm}} \ \left(\frac{16}{A}\right)^{\frac{1}{2}} \left(\frac{10^{3} \ \rm K}{T}\right)^{\frac{1}{2}} \\ \nonumber & \simeq 10^{-9}  \ {\rm{cm}} \ \left(\frac{130}{A}\right)^{\frac{1}{2}} \left(\frac{177 \ \rm K}{T}\right)^{\frac{1}{2}},
     \label{eq:dEcoh}
     \end{align}
     
     where in the upper line we have normalized the expression for $^{16}$O, which is the most abundant element in mass and concentration in the Earth's crust and mantle, and a temperature of $\sim 10^{3}$ K. The expression in the lower line, on the other hand, has been normalized to $^{130}$Xe and a temperature $\sim 177 \ \rm K$, corresponding to the parameters of the \acro{XENON-1T} experiment, which applies to Section~\ref{sec:migdal}.
     
\end{itemize}

\subsection{Single-Particle Excitations}
\label{sec:nucdmint2}

In addition to quasi-coherent composite-nucleus scattering, accelerated nuclei in the interior of the composite may lose energy by exciting individual dark matter constituents above their Fermi level. We consider the dark matter constituents to form a non-interacting Fermi gas of particles with effective mass $m_*$, with $m_* \ll p_{F}$, when the composite state is sufficiently massive, $cf.$ Eqs.~\eqref{eq:sat-comp-meff} and \eqref{eq:sat-comp-chem}. Since the dark matter constituents are ultra-relativistic, we must then carefully analyse both the phase space available for scattering and the relativistic kinematics of the collisions. In \cite{Joglekar:2019vzy,Joglekar:2020liw}, a Lorentz-invariant formalism was developed to compute the dark matter capture efficiency of relativistic degenerate electrons in neutron stars. Such formalism accounts for both relativistic kinematics and Pauli blocking of the targets. Here, we adapt those results to estimate the scattering rate and the subsequent energy loss of nuclei inside the composite from exciting constituents above the Fermi level. To begin, we consider the scattering rate of a nucleus against dark matter constituents in the composite rest frame \cite{Joglekar:2020liw},

\begin{equation}
\Gamma_{NX \rightarrow NX^{*}} = n_{X} \int_{p_{\rm min}}^{p_{F}} \frac{dp \ p^{2}}{V_{F}} \int d\varphi \ d(\cos\theta) \int d\alpha \ d(\cos\psi) \ {\rm \Theta}\left(\Delta E + p - p_{F}\right) \ \tilde{v} \left(\frac{d\sigma}{d\Omega}\right)_{(CM)},
\label{eq:spgamma}
\end{equation}
where  $V_{F}=4\pi p_{F}^{3}/3$ is the occupied phase space volume of the target dark matter particles, $n_{X}= p_{F}^{3}/3\pi^{2}$ is the dark matter number density, $(d\sigma/d\Omega)_{(CM)}$ is the dark matter-nucleus differential cross section in the centre-of-momentum frame, and

\begin{equation}
\tilde{v}=\left(\frac{(p_{\mu} k^{\mu})^{2}-m_{*}^{2}m_{N}^{2}}{(p^{2}+m_{*}^{2})(m_{N}^{2}+k^{2})}\right)^{1/2}
\label{eq:spmoller}
\end{equation}
is the Moller velocity \cite{landau2013classical,Cannoni:2016hro}, with $p^{\mu}$ and $k^{\mu}$ respectively being the four-momentum of the dark matter particle and the nucleus in the composite rest frame. The latter is determined from Eq.~\eqref{eq:accelsr3}. The Moller velocity relates differential cross sections in different frames for non-collinear scattering processes. In this case, the frames considered are the composite rest frame, where the Fermi surface is spherical, and the centre-of-momentum frame, where the collision kinematics are more easily analysed. The scattering rate given by Eq.~\eqref{eq:spgamma} integrates over the target Fermi sphere, and accounts for the potentially large boost between both frames for relativistic targets. The first two integrals correspond to integrating the differential cross section over the azimuthal and polar scattering angles $(\alpha,\psi)$ in the center-of-momentum frame. The integrals that follow correspond to integrating over the azimuthal and polar angles $(\varphi,\theta)$ on the Fermi sphere (the integral over $\varphi$ yields a $2\pi$ factor). These angles determine the momentum direction of the dark matter targets in the composite rest frame. The final integral to the left corresponds to integrating over the momentum magnitude $p$ of the dark matter constituents, also in the composite rest frame. Note that it runs from some $p=p_{\rm min}$ set by the Pauli blocking condition, $i.e.$ only those dark matter particles occupying states sufficiently close to the Fermi surface will be excited. This condition is enforced by the Heaviside step function, where $\Delta E$ is the energy deposited by the nucleus in the composite frame, see below. 

The differential nucleus-dark matter scattering cross section, in the centre-of-momentum frame, is determined by the t-channel matrix element of the Yukawa interaction, suitably averaged over spins,
\begin{equation}
\left(\frac{d\sigma}{d\Omega}\right)_{(CM)}=\frac{\langle|\mathcal{M}_{t}|^{2} \rangle}{64 \pi^{2} s}=\frac{3 A^{2} g_{n}^{2}g_{X}^{2}}{16\pi^{2}} \frac{t}{(t+m_{\phi}^{2})^{2}}.
\label{eq:spcs}
\end{equation}
In the above expression, we have neglected the effective mass $m_{*} \ll p_{F}$ of the dark matter constituents. We do not include the Helm form factor as it evaluates to unity for the momentum exchanges involved here. The Mandelstam variables are $s=E_{cm}^2$ and $t=4k_{cm}^{2} \sin^{2} \left(\psi/2\right)$, where $k_{cm}=|\mathbf{k}_{cm}|$ and $E_{cm}$ respectively are the 3-momentum magnitude of the nucleus and total energy in the center-of-momentum frame. 

The energy transfer $\Delta E$ in the composite rest frame is \cite{Joglekar:2020liw}

\begin{equation}
\Delta E = \gamma \beta k_{cm} \left(\cos \delta (1-\cos \psi)-|\sin \delta| \cos \alpha \sin \psi \right),
\label{eq:spDE}
\end{equation}
where $\gamma$ and $\beta=|\boldsymbol{\beta}|$ are the boost parameters relating the composite rest frame to the centre-of-momentum frame of the nucleus-dark matter system,

\begin{equation}
\boldsymbol{\beta}=\frac{\mathbf{p}+\mathbf{k}}{p+m_{N}} \ \ ,\  \ \gamma=\frac{1}{\sqrt{1-\beta^{2}}}.
\end{equation}

The function $\cos\delta$ in Eq.~\eqref{eq:spDE} is

\begin{equation}
 \cos\delta = \frac{p k^{2}-m_{N}p^{2}+(p-m_{N})\mathbf{p}\cdot \mathbf{k}}{(p+m_{N})\beta E_{cm} k_{cm}},
 \label{eq:spcosdelta}
\end{equation}
where this quantity determines the kinematic suppression of the scattering angles $(\alpha,\psi)$. In short, when $\cos\delta>0$ the integration over $(\alpha,\psi)$ is unconstrained. However, if $\cos\delta<0$ there is a maximum scattering angle $\psi_{\rm max}$ for which the energy transfer $\Delta E$ is positive and satisfies the Pauli blocking condition, see \cite{Joglekar:2020liw} for details. In the non-relativistic limit considered here (see Appendix~\ref{sec:app-c}), $\cos\delta \simeq -1$, resulting in a substantial suppression of the available scattering phase space. The maximum scattering angle is \cite{Joglekar:2020liw},

\begin{equation}
    \tan^2 \left(\frac{\psi_{\rm max}}{2}\right) = \frac{k^2 p^2 \sin^2 \theta \left(m_N (m_N + 2p) -2 p k \cos\theta\right)}{p k^2 - m_N p^2 (p-m_N) p k \cos\theta}\cos^2\alpha.
\end{equation}
For the parameter space considered, $\psi_{\rm max} \ll 1$ which substantially simplifies integration of Eq.~\eqref{eq:spgamma}. The maximum energy transfer occurs for $\alpha = \pi$ and $\cos \psi = - \cos \delta$, and is given by

\begin{equation}
    \Delta E_{\rm max} = \gamma \beta k_{cm} \left(\cos\delta + 1\right)
\end{equation}
We remark that $\Delta E_{\rm max}$ depends on $\theta$, $i.e.$ it is determined by the momentum direction of the dark matter particle in the composite rest frame. In Appendix~\ref{sec:app-c}, we provide the non-relativistic expressions of all the kinematic quantities listed above, which are obtained by replacing $\mathbf{k} \simeq m_N \mathbf{v}_N$, and Taylor-expanding in $v_N$. In order to analytically integrate  Eq.~\eqref{eq:spgamma}, we assume a maximal energy loss independent of the scattering directions, which conservatively maximizes the Fermi shell where scattering is kinematically permitted. The details of the computation can also be found in Appendix~\ref{sec:app-c}. With this assumption, we find the rate at which a non-relativistic nucleus scatters against composite constituents is,

\begin{equation}
\Gamma_{NX \rightarrow NX^{*}}\simeq \frac{3 A^{2} g_{n}^{2}g_{X}^{2} m_N^4 (m_N+2p_F)v_N^6}{40\pi p_F^4}
\end{equation}

This scattering rate is strongly suppressed by an $\mathcal{O}(v_N^6)$ factor at the lowest order in the nuclear velocity, with $v_N \lesssim 10^{-2}$ for the most energetic signatures of this model. Following our conservative assumption of maximal energy loss, we estimate the energy loss rate by multiplying the above expression by $\Delta E_{\rm max}$,

\begin{equation}
    \dot{E}_{NX \rightarrow NX^{*}} \simeq \Delta E_{\rm max} \ \Gamma_{NX \rightarrow NX^{*}} \simeq \frac{3 A^{2} g_{n}^{2}g_{X}^{2} m_N^5 (m_N+2p_F)v_N^8}{80\pi p_F^4},
    \label{eq:spEdot}
\end{equation}
Given the $\mathcal{O}(v_N^8)$ suppression, this result demonstrates that, due to the relativistic kinematics and degeneracy considerations, scattering between the two particle species is extremely suppressed. As we discuss in Appendix~\ref{sec:app-c}, this expression largely overestimates the scattering phase space available and, in practice, this energy loss rate should be much smaller. 
The total stopping power from this inelastic process in this approximation is then

\begin{equation}
    \left(\frac{dE}{dx}\right)_{NX \rightarrow NX^{*}}\simeq n_N \left(\frac{4 \pi R_X^3}{3}\right) \dot{E}_{NX \rightarrow NX^{*}} 
    \label{eq:spdEdx}
\end{equation}

We conclude that the stopping power from inelastic nuclear-constituent interactions is negligible for the strongly-bound composites we have studied. We remark that the above computation implicitly assumes that an excitation created by a nucleus is uncorrelated with excitations created by other nuclei. In general, the up-scattered dark matter particle, as well as the hole below the Fermi surface that is created, will have complicated dynamics since we are considering dark matter particles with self-interactions. A detailed study of how these excitations evolve, and how the composite decays back to its ground state, is beyond the scope of this paper and left for future study. However, from Fermi liquid theory, we expect the up-scattered dark matter particle to have a rapid decay rate proportional to the binding energy $\tau^{-1}\sim (m_{X}-\bar{m}_{X})$, much greater than the rate $\Gamma_{X N}$ at the which dark matter particles are up-scattered by nuclei. Therefore, we expect the above estimate for the total energy loss rate to hold.

\section{Conclusions}
\label{sec:concl}

We have explored a new effect recently identified in \cite{Acevedo:2020avd}, whereby the Yukawa potential that binds together composite DM can accelerate Standard Model particles at the composite periphery. The asymmetric DM composites we focused on consist of a dark matter fermion coupled to a real scalar field, which provides the attractive force to form bound states. Such large bound states form in the early universe when smaller bound states fuse into bigger ones, followed by the emission of scalar mediators, so long as the binding energy per dark matter particle is comparable to its mass, $i.e.$ the strong binding limit. This synthesis process may also include a subsequent dark matter dilution stage that depletes any excess abundance, permitting the formation of bound states with constituent masses ranging from MeV - EeV. 

A Yukawa interaction between nuclei and the scalar composite binding field results in rich phenomenology when the composite sizes exceed $\gtrsim{ \rm nm}$ scales. Since the binding scalar field takes on values $\propto m_X$, even for a minuscule DM-SM Yukawa coupling, nuclei can substantially accelerate at the composite boundary. The composite DM signatures we have identified depend on the coupling strength and the sign of the Yukawa term, and include heat dissipation, collisional ionization, thermal Bremsstrahlung, nuclear recoils, ionization via the Migdal effect, and even thermonuclear reactions. New aspects of this study include a detailed analysis of the different scattering processes between the nuclei and the dark matter, an expanded discussion on white dwarf explosions induced by composite transit, implications for planetary heating, and direct detection prospects at dark matter experiments from the Migdal effect.

Due to the large composite size, the small dark matter-nucleon coupling, and mediator screening effects, coherent scattering of the composite with nuclei, as well as the excitation of low-lying collective modes, is largely suppressed. On the other hand, we have also considered single particle excitations whereby a nucleus scatters against individual constituents, exciting them above the Fermi level. This analysis is complicated by the fact that, for large composite states, the dark matter constituents are highly-degenerate and relativistic, resulting in a tiny phase space region where scattering can proceed. Because of such kinematic considerations, we also find that nuclei do not scatter frequently with the constituents. In short, nuclei losing energy by interacting with the dark matter is negligible. By contrast, nuclei also lose energy via heat dissipation in the form of conduction and radiation, and these processes are the main contribution to the stopping power. The exact energy loss rate depends on the coupling strength as well as the specific material the composite state is transiting. 

We have studied in further detail how large composite states can ignite a carbon-oxygen white dwarf by simply passing through its stellar core. Nuclei accelerated in the composite interior can fuse in the thermonuclear regime, provided their temperature reaches $\sim \mev$, leading to a nuclear runaway if the composite is sufficiently large. The critical ignition size of the composited was determined by requiring that nuclear energy release overcomes heat dissipation rate in the white dwarf material. We have found that composites with radii $R_X \gtrsim 10^{-2} \ \rm \mu m$ and masses $M_X \gtrsim 10^{30} ~\gev$ are capable of reaching and igniting the core of massive white dwarfs. Based on the survival of these stellar objects on $\sim$Gyr timescales, we have placed bounds on the dark matter-nucleon coupling in the composite mass range $10^{30} \ \gev \lesssim M_X \lesssim 10^{42} \ \gev$, where the upper mass limit is determined from requiring at least one encounter with a composite over a $\sim \rm Gyr$ timescale for a dark matter density of order $\rho_X \simeq 0.4 \ \gev \ \rm cm^{-3}$. We have discussed as well the implications of this model for planetary heating, and find that although a substantial fraction of composites could be captured by the Earth, composite capture alone cannot account for the entire heat flow observed, even if they induce exothermic fusion reactions in the mantle.

Finally, we have analysed for the first time the direct detection prospects for DM composites with very weak nuclear couplings at noble liquid experiments by considering the Migdal effect, which in this case consists of the excitation and ionization of electrons from the impulsive motion of an atomic nucleus as it accelerates at the boundary of a DM composite. For \acro{Xenon-1T}, this effect is particularly potent at putting bounds on DM composite couplings to nuclei, given the low electron background of this experiment, combined with the high ionization probabilities of the outer shell electrons in xenon atoms. We have computed the expected number of electron recoils using the exposure of \acro{XENON-1T}'s first dark matter search, and used it to place bounds on the dark matter-nucleon coupling, in the mass range $10^{12} \ \gev \lesssim M_X \lesssim 10^{17} \ \gev$, where these limits arise from requiring the composite to be strongly bound and be larger than the size of a xenon nucleus. The constraints we placed are as low as $g_n \lesssim 10^{-17}$, and lie well below existing bounds from stellar cooling arguments and previous dark matter searches based on the observation $\sim \kev$ nuclear recoils. We have also found a significant region of parameter space where composites are large enough to produce ionization tracks. At present, these coupling values cannot be ruled out as this signature would require a dedicated analysis of the experiment's data, which accounts for multiple Migdal electron ionization events over the course of composite DM's transit through the liquid xenon. We have left the study of this regime, as well as certain aspects of weakly-bound DM composites, for future work.  

\section*{Acknowledgements}

We thank Nirmal Raj for useful discussions. We thank the anonymous referee for constructive comments on our manuscript. The work of JA, JB, AG is supported by the Natural Sciences and Engineering Research Council of Canada (NSERC). Research at Perimeter Institute is supported in part by the Government of Canada through the Department of Innovation, Science and Economic Development Canada and by the Province of Ontario through the Ministry of Colleges and Universities.

\appendix

\section{Energy Dissipation and Composite Stopping}
\label{sec:app-a}

In this Appendix, we detail how a dark matter composite slows down due to nuclei losing kinetic energy from various scattering processes occurring in their interior. This can be understood from considering momentum and energy conservation: if a nucleus loses kinetic energy while transiting the composite, then when it exits the composite its momentum will be more aligned with the direction of motion of the composite. The net result is that the dark matter composite slows down. Below we confirm that the decrease in DM composite kinetic energy matches the kinetic energy transferred to the nucleus inside the composite, in the non-relativistic limit considered in this work. This treatment will be specific to the case that nuclei are accelerated by an attractive potential inside the DM composite -- in the case of a repulsive potential, standard two body kinematics apply. 

Let us consider a single nucleus entering the composite volume while the composite moves through a given stellar (or planetary) medium with an initial speed $v_i$ in the stellar rest frame. In what follows, we assume the center-of-mass (CM) frame for this system is equivalent to the composite rest frame, since we are always in the limit $m_N \ll M_X$. Upon acceleration from the composite potential, energy conservation in the CM frame reads

\begin{equation}
   \frac{1}{2}m_N v_i^2 = \frac{1}{2}m_N v_N^2 + \varphi,
\end{equation}
where $v_N$ is the velocity of the nucleus after accelerating to the final kinetic energy and $\varphi<0$ is the potential energy, with $|\varphi| \ll m_N$ in the non-relativistic limit, $cf.$ Eq.~\eqref{eq:accelsr1}. In the stellar rest frame, this velocity is $(v_N)_{\rm sta}=v_i - \left(v_i - 2 \varphi /m_N\right)^{1/2}$, and momentum conservation imposes

\begin{equation}
    M_X v_i = M_X (v_X)_{\rm sta} + m_N (v_N)_{\rm sta}.
\end{equation}
This yields a velocity for the composite after the nucleus has accelerated given by $(v_X)_{\rm sta} = v_i - (m_N/M_X) \left(v_i - \left(v_i - 2 \varphi /m_N\right)^{1/2} \right)$. Note that since $m_N \ll M_X$, the composite barely recoils in the process. 

As discussed in the main text, the nucleus can undergo several inelastic processes while inside the composite state that reduce its kinetic energy by an amount we denote here by $\delta{E}<0$. Furthermore, we assume $|\delta E| \ll |\varphi|$, which is the case throughout this work. If $\delta{E} \sim \varphi$ then the nucleus would remain bound to the composite and the analysis below would be different. Upon deceleration, energy conservation in the CM frame now is,

\begin{equation}
   \frac{1}{2} m_N (v_N')^2 + \varphi = \frac{1}{2} m_N (v_N'')^2,
\end{equation}
where the velocity $v_N'$ now is related to $v_N$ above via $\delta{E}=(m_N/2)(v_N'^2 - v_N^2)$. Then, $v_N' = - \left(v_N^2 + 2 \delta E / m_N\right)^{1/2} = -\left(v_i^2 + 2(\delta E - \varphi)/m_N\right)^{1/2}$, and we solve for $v_N''$ in the above equation to obtain $v_N''= - \left(v_i^2 + 2\delta E/m_N\right)^{1/2}$. Boosting back to the stellar rest frame, conservation of momentum now requires

\begin{equation}
    M_X (v_X)_{\rm sta} + m_N (v_N')_{\rm sta} = M_X (v_X')_{\rm sta} + m_N (v_N'')_{\rm sta},
\end{equation}
where the unknown velocity above is $(v_X')_{\rm sta}$. The two velocities computed above in the stellar rest frame are $(v_N')_{\rm sta} = v_i -\left(v_i^2 + 2(\delta E - \varphi)/m_N\right)^{1/2}$ and $(v_N'')_{\rm sta} = v_i - \left(v_i^2 + 2\delta E/m_N\right)^{1/2}$. Solving for $(v_X')_{\rm sta}$ above yields,

\begin{equation}
    (v_X')_{\rm sta} - v_i = \frac{m_N}{M_X} \left(v_i - \left(v_i + \frac{2 (\delta E - \varphi)}{m_N}\right)^{1/2} - v_i + \left(v_i^2 - \frac{2 \varphi}{m_N}\right)^{1/2} - v_i + \left(v_i^2 + \frac{2\delta E}{m_N}\right)^{1/2}\right)
\end{equation}
In the limit $|\delta E| \ll |\varphi|$, the first four terms on the right hand side cancel out. Similarly, we can expand $\left(v_i^2 + 2 \delta E /m_N\right)^{1/2} = v_i \left(1 + 2 \delta E /m_N v_i^2\right)^{1/2} \simeq v_i (1+\delta E / m_N v_i^2)$ in the limit $|\delta E| \ll m_N v_i^2$. Therefore,

\begin{equation}
    (v_X')_{\rm sta} - v_i \simeq \frac{\delta E}{M_X v_i}.
\end{equation}
Squaring this expression and multiplying by $M_X/2$ yields,

\begin{equation}
    \frac{1}{2} M_X (v_X')_{\rm sta}^2 = \frac{1}{2} M_X \left(v_i + \frac{\delta E}{M_X v_i}\right)^2 \simeq \frac{1}{2} M_X \left(v_i^2 + \frac{2 \delta E}{M_X}\right)
\end{equation}

Thus, in the stellar rest frame, the kinetic energy of the composite is reduced by

\begin{equation}
    \frac{1}{2} M_X (v_X')_{\rm sta}^2 - \frac{1}{2} M_X v_i^2 \simeq \delta E < 0.
\end{equation}
This result implies that for nuclei traveling through an attractive potential inside a DM composite, any nuclear energy dissipation processes occurring in the composite interior, such as heat conduction, radiation or endothermic reactions, will continuously decrease the DM composite kinetic energy, as viewed in the stellar rest frame. On the other hand, if there are processes that heat the nuclear matter, such as exothermic reactions, the stopping power would be subsequently reduced.

\section{Collective Modes}
\label{sec:app-b}

Here we address nuclear scattering and excitation of collective modes of constituent particles comprising the dark matter composite. A full treatment of composite collective modes is out of the scope of this work, as there are many possible modes; namely surface and compressional modes, rotational modes, spin-waves, and so forth. The excitation rate of the different collective modes, for a given momentum transfer $\mathbf{q}$ and energy transfer $\omega$, will be encoded in an appropriate response function $S_X(\mathbf{q},\omega)$ which, to our knowledge, has not been yet computed for this dark matter model. Ref.~\cite{Hardy:2015boa} studied the lowest-lying vibrational surface modes for such composite states, which have an excitation gap of order

\begin{equation}
 \Delta E_{\rm surf} \simeq \left(\frac{\epsilon_{\rm surf} \bar{m}_{X}}{N_{X}}\right)^{\frac{1}{2}} \simeq 2.2 \ {\rm eV} \left(\frac{m_X}{\rm TeV}\right)^\frac{1}{2} \left(\frac{\bar{m}_X}{5 \ \gev}\right)^\frac{1}{2} \left( \frac{N_X}{10^{20}}\right)^{-\frac{1}{2}},
 \label{eq:app-b-dE}
\end{equation}
where $\epsilon_{\rm surf} \simeq m_X - \bar{m}_X \simeq m_X$ is a constant related to the surface energy of the composite state \cite{Gresham:2017zqi}. Although in terms of the energy required these modes could easily be excited, the cross section for creating a single phonon with angular momentum number $l$ is highly suppressed due to the large composite size and mediator screening. The cross section for such inelastic process can be expressed as

\begin{equation}
\left(\frac{d\sigma}{dq}\right)_{0 \rightarrow 1_{l}} \simeq A^2 N_{X}^{2} f^2(\Lambda) \bar{\sigma}_{0} \left(\frac{q}{2 m_{N}^{2}v_{X}^{2}}\right) |F_{\phi}(q)|^{2} |F_{\rm surf}^{(l)}(qR_{X})|^{2},
\end{equation}
where, as before, we include the mediator mass correction due to screening in $\bar{\sigma}_0$, $cf.$ Eq.~\eqref{eq:csref1}, as well as the ratio between the scatterer wavefunction spread $\Lambda$ and the composite size $R_X$, given by Eq.~\eqref{eq:blob}. The cross section for exciting surface modes has an additional form factor 

\begin{equation}
F_{\rm surf}^{(l)}(qR_{X})=\epsilon_{l} \ (2l+1)^{1/2}  j_{l}(qR_{X}).
\end{equation}
where the quantity $\epsilon_{l}$ is the amplitude of the mode, which scales as,

\begin{equation}
\epsilon_{l} \propto {m_X^{-1/4} \bar{m}_X^{-3/2} R_X^{-7/4}} \simeq 10^{-14} \ \left(\frac{m_X}{\rm TeV}\right)^{-\frac{1}{4}} \left(\frac{\bar{m}_X}{5 \ \gev}\right)^{-\frac{3}{2}} \left(\frac{R_X}{\rm nm}\right)^{-\frac{7}{4}}.
\end{equation}

The creation of multiple vibration quanta in a given mode $l$ result in extra powers of $\epsilon_l$, and therefore further suppression of the cross section. Computing the stopping power, $cf.$ Section~\ref{sec:nucdmint1}, we have verified that the energy loss from this inelastic process is negligible. Similarly, we expect other collective modes to be suppressed from the same screening effects as well as having potentially higher excitation gaps.

\section{Scattering Rate against Dark Matter Constituents}
\label{sec:app-c}

In this Appendix, we provide details of the phase space integration that yields the scattering and energy loss rate from nuclei scattering against single dark matter constituents, Eq.~\eqref{eq:spgamma}, in the limit that the nucleus is non-relativistic, $i.e.$ $v_N \ll 1$. In such case, all the kinematic quantities are considerably simplified. In the composite rest frame, the four-momentum of the nucleus is $k^{\mu}\simeq (m_{N},\mathbf{k})$, with $\mathbf{k} \simeq m_N \mathbf{v}_N$. A dark matter constituent, on the other hand, will be ultra-relativistic, $i.e.$ $m_{*} \ll p_{F}$, with a four-momentum $p^{\mu}\simeq (p,\mathbf{p})$. We neglect the effective mass of the constituent $m_{*}$ as the nucleus will only scatter off dark matter constituents with momenta very close to the Fermi surface.

The relevant kinematic variables are the Moller velocity $\tilde{v}$, center-of-momentum energy $E_{cm}$ and the nucleus 3-momentum magnitude $k_{cm}$, the boost parameter $\mathbf \beta$ relating the center-of-momentum frame to the composite rest frame, $\cos\delta$ given by Eq.~\eqref{eq:spcosdelta}, the maximum energy transfer $\Delta E_{max}$, and the maximum scattering angle $\psi_{max}$. To lowest order in the nuclear speed $v_N$, these quantities read

\begin{equation}
\tilde{v} \simeq 1-v_N \cos\theta,
\end{equation}

\begin{equation}
    E_{cm}^2 \simeq m_{N}^{2}+2m_{N}p(1-v_N\cos\theta)
\end{equation}

\begin{equation}
    k_{cm}^2 \simeq \frac{m_N p^2}{m_N+2p} - \frac{2 m_N p^2(m_N+p)v_N \cos \theta}{(m_N+2p)^2} 
\end{equation}

\begin{equation}
    \beta \simeq \frac{p}{m_N + p} + \frac{m_N v_N \cos \theta}{m_N+p}
\end{equation}

\begin{equation}
 \cos\delta \simeq -1 + \frac{m_N (m_N+2p)v_N^2\sin^2\theta}{2p^2},
\end{equation}

\begin{equation}
    \Delta{E_{max}} \simeq \frac{1}{2} m_N v_N^2 \sin^2 \theta
    \label{eq:app-c-dEmax}
\end{equation}

\begin{equation}
    \psi_{max} \simeq \frac{\left(m_N (m_N+2p)\right)^{1/2} v_N \cos \alpha}{p}
\end{equation}
Note that $\cos \delta \simeq -1$ in the non-relativistic limit, and therefore the scattering angle is bounded by $\psi \lesssim \psi_{max}$. This maximum angle is in turn $\psi_{max} \ll 1$, and so the integration of the differential cross section, $cf.$ Eq.~\eqref{eq:spcs}, is considerably simplified. We can approximate $\sin^2 \psi/2 \simeq \psi^2/4$, so the Mandelstam variable is $t \simeq k_{cm}^{2} \psi^{2}$. Furthermore, we include the correction to the mediator mass from the dense dark matter medium, given by $\delta m_{\phi}^{2}\simeq g_{X}^{2} \langle \bar{X} X \rangle / \bar{m}_{X} = \left(4 \pi \alpha_X p_F^4 \right)^{1/2}$. With these considerations, the cross section reads

\begin{equation}
 \left(\frac{d\sigma}{d\Omega}\right)_{\rm (CM)} \simeq \frac{3 A^{2} g_{n}^{2}g_{X}^{2}}{16\pi^{2}} \frac{k_{cm}^{2}\psi^{2}}{(k_{cm}^{2}\psi^{2}+\tilde{m}_{\phi}^{2})^{2}},
\end{equation}
where $\tilde{m}_\phi^2 = m_\phi^2 + \delta m_\phi^2$. Approximating $d(\cos\psi)=(\sin\psi) d\psi \simeq \psi d\psi$ as well in this small angle limit, we can analytically integrate the cross section over the scattering angle $\psi$,

\begin{equation}
    \int_{0}^{\psi_{max}} d(\cos\psi) \ \tilde{v} \left(\frac{d\sigma}{d\Omega}\right)_{\rm (CM)} = \frac{3 A^{2} g_{n}^{2}g_{X}^{2}}{32\pi^{2} k_{cm}^{2}} \left[\log\left(1+x^2\right)-\frac{x^2}{1+x^2}\right] (1-v_N\cos\theta),
\end{equation}
Here we have defined $x = k_{cm}\psi_{max}/\tilde{m}_{\phi}$. This ratio is $x \ll 1$ in the parameter space considered here, and so we expand both functions above using $\log(1+x^{2}) \simeq x^{2} -x^{4}/2 + \mathcal{O}(x^{6})$ and $x^{2} (1+x^{2})^{-1} \simeq x^{2} - x^{4} + \mathcal{O}(x^{6})$. The first terms in each series mutually cancel, and so the lowest-order contribution is $\mathcal{O}(x^{4})$. The integrated cross section is then simplified to,

\begin{equation}
\int_{0}^{\psi_{max}} d(\cos\psi) \ \tilde{v} \left(\frac{d\sigma}{d\Omega}\right)_{\rm (CM)} \simeq \frac{3 A^{2} g_{n}^{2}g_{X}^{2}}{64\pi^{2}} \frac{k_{cm}^{2} \psi_{max}^{4}}{\tilde{m}_\phi^{4}} (1-v_N\cos\theta).
\end{equation}

Since $\psi_{max}^4 \sim v_N^4$, we only need to consider the lowest order term in $k_{cm}^2$ and the Moller velocity $\tilde{v}$. In order to perform analytically the remaining integrals in Eq.~\eqref{eq:spgamma}, we assume that the energy transfer is always maximal. This estimate is highly conservative, since in practice the energy transfer $\Delta E$ is peaked around a specific $\psi^{*} < \psi_{max}$ with $\alpha=\pi$ and $\theta=\pi/2$, while any other angle values $\alpha,\psi,\theta$ yield an energy exchange smaller by several orders of magnitude. The lower integration $p_{min}$ in Eq.~\eqref{eq:spgamma} is then independent of the angular integration variables, and set by the condition

\begin{equation}
p_{min} \simeq p_{F}-\frac{1}{2} m_{N} v_N^{2}.
\end{equation}
Thus, Eq.~\eqref{eq:spgamma} splits into

\begin{align}
    \int^{p_F}_{p_{min}} & dp \ p^2 \int_{0}^\pi d(\cos\theta) \int_0^{2\pi} d\alpha \left(\frac{3 A^{2} g_{n}^{2}g_{X}^{2}}{64\pi^{2}} \frac{k_{cm}^{2} \psi_{max}^{4}}{\tilde{m}_\phi^{4}}\right) \simeq \\ \nonumber & \simeq \left(\frac{3 A^{2} g_{n}^{2}g_{X}^{2} m_N^3 v_N^4}{64\pi^{2}p_F^4}\right) \times \left(\int^{p_F}_{p_{min}} dp \ (m_N+2p)\right) \times \left(\int_0^{\pi} d(\cos\theta) \int_0^{2\pi} \cos^4 \alpha \right).
\end{align}
The remaining angular integration yields a factor $8 \pi /5$, whereas integration of the momentum magnitude $p$ yields a factor $\sim (m_N + 2p_F)(p_F - p_{min})$, in the limit that $p_{min} \ll p_F$. The resulting scattering rate is therefore,

\begin{equation}
\Gamma_{NX \rightarrow NX^{*}}\simeq \frac{3 A^{2} g_{n}^{2}g_{X}^{2} m_N^4 (m_N+2p_F)v_N^6}{40\pi p_F^4}.
\end{equation}
Multiplying the above rate by the maximum energy transfer Eq.~\eqref{eq:app-c-dEmax}, we recover Eq.~\eqref{eq:spEdot}. This expression is usefully valid for $m_N$ either greater or smaller than $p_F$, and demonstrates that even with the assumption of maximal energy loss in any scattering direction, the excitation of dark matter constituents above the Fermi level in highly degenerate ``saturated'' composites, is highly infrequent due to a resulting $\mathcal {O}(v_N^6)$ velocity suppression of the scattering rate. It follows that individual dark matter particle excitations are not a relevant energy loss channel for nuclei traveling through the composite interior. 

\bibliographystyle{JHEP}

\bibliography{cdm}

\providecommand{\href}[2]{#2}\begingroup\raggedright\begin{thebibliography}{100}

\bibitem{Wise:2014ola}
M.~B. Wise and Y.~Zhang, \emph{{Yukawa Bound States of a Large Number of
  Fermions}}, \href{http://dx.doi.org/10.1007/JHEP02(2015)023}{\emph{JHEP} {\bf
  02} (2015) 023}, [\href{http://arxiv.org/abs/1411.1772}{{\tt 1411.1772}}].

\bibitem{Wise:2014jva}
M.~B. Wise and Y.~Zhang, \emph{{Stable Bound States of Asymmetric Dark
  Matter}}, \href{http://dx.doi.org/10.1103/PhysRevD.90.055030}{\emph{Phys.
  Rev. D} {\bf 90} (2014) 055030}, [\href{http://arxiv.org/abs/1407.4121}{{\tt
  1407.4121}}].

\bibitem{Gresham:2017cvl}
M.~I. Gresham, H.~K. Lou and K.~M. Zurek, \emph{{Early Universe synthesis of
  asymmetric dark matter nuggets}},
  \href{http://dx.doi.org/10.1103/PhysRevD.97.036003}{\emph{Phys. Rev. D} {\bf
  97} (2018) 036003}, [\href{http://arxiv.org/abs/1707.02316}{{\tt
  1707.02316}}].

\bibitem{Gresham:2017zqi}
M.~I. Gresham, H.~K. Lou and K.~M. Zurek, \emph{{Nuclear Structure of Bound
  States of Asymmetric Dark Matter}},
  \href{http://dx.doi.org/10.1103/PhysRevD.96.096012}{\emph{Phys. Rev. D} {\bf
  96} (2017) 096012}, [\href{http://arxiv.org/abs/1707.02313}{{\tt
  1707.02313}}].

\bibitem{Acevedo:2020avd}
J.~F. Acevedo, J.~Bramante and A.~Goodman, \emph{{Nuclear fusion inside dark
  matter}}, \href{http://dx.doi.org/10.1103/PhysRevD.103.123022}{\emph{Phys.
  Rev. D} {\bf 103} (2021) 123022},
  [\href{http://arxiv.org/abs/2012.10998}{{\tt 2012.10998}}].

\bibitem{Hardy:2014mqa}
E.~Hardy, R.~Lasenby, J.~March-Russell and S.~M. West, \emph{{Big Bang
  Synthesis of Nuclear Dark Matter}},
  \href{http://dx.doi.org/10.1007/JHEP06(2015)011}{\emph{JHEP} {\bf 06} (2015)
  011}, [\href{http://arxiv.org/abs/1411.3739}{{\tt 1411.3739}}].

\bibitem{Hardy:2015boa}
E.~Hardy, R.~Lasenby, J.~March-Russell and S.~M. West, \emph{{Signatures of
  Large Composite Dark Matter States}},
  \href{http://dx.doi.org/10.1007/JHEP07(2015)133}{\emph{JHEP} {\bf 07} (2015)
  133}, [\href{http://arxiv.org/abs/1504.05419}{{\tt 1504.05419}}].

\bibitem{Gresham:2018anj}
M.~I. Gresham, H.~K. Lou and K.~M. Zurek, \emph{{Astrophysical Signatures of
  Asymmetric Dark Matter Bound States}},
  \href{http://dx.doi.org/10.1103/PhysRevD.98.096001}{\emph{Phys. Rev. D} {\bf
  98} (2018) 096001}, [\href{http://arxiv.org/abs/1805.04512}{{\tt
  1805.04512}}].

\bibitem{Petraki:2013wwa}
K.~Petraki and R.~R. Volkas, \emph{{Review of asymmetric dark matter}},
  \href{http://dx.doi.org/10.1142/S0217751X13300287}{\emph{Int. J. Mod. Phys.
  A} {\bf 28} (2013) 1330028}, [\href{http://arxiv.org/abs/1305.4939}{{\tt
  1305.4939}}].

\bibitem{Zurek:2013wia}
K.~M. Zurek, \emph{{Asymmetric Dark Matter: Theories, Signatures, and
  Constraints}},
  \href{http://dx.doi.org/10.1016/j.physrep.2013.12.001}{\emph{Phys. Rept.}
  {\bf 537} (2014) 91--121}, [\href{http://arxiv.org/abs/1308.0338}{{\tt
  1308.0338}}].

\bibitem{Walecka:1995mi}
J.~Walecka, \emph{{Theoretical nuclear and subnuclear physics}}, vol.~16.
\newblock 1995.

\bibitem{Bramante:2017obj}
J.~Bramante and J.~Unwin, \emph{{Superheavy Thermal Dark Matter and Primordial
  Asymmetries}}, \href{http://dx.doi.org/10.1007/JHEP02(2017)119}{\emph{JHEP}
  {\bf 02} (2017) 119}, [\href{http://arxiv.org/abs/1701.05859}{{\tt
  1701.05859}}].

\bibitem{Affleck:1984fy}
I.~Affleck and M.~Dine, \emph{{A New Mechanism for Baryogenesis}},
  \href{http://dx.doi.org/10.1016/0550-3213(85)90021-5}{\emph{Nucl. Phys.} {\bf
  B249} (1985) 361--380}.

\bibitem{Dine:1995kz}
M.~Dine, L.~Randall and S.~D. Thomas, \emph{{Baryogenesis from flat directions
  of the supersymmetric standard model}},
  \href{http://dx.doi.org/10.1016/0550-3213(95)00538-2}{\emph{Nucl. Phys.} {\bf
  B458} (1996) 291--326}, [\href{http://arxiv.org/abs/hep-ph/9507453}{{\tt
  hep-ph/9507453}}].

\bibitem{Ebadi:2021cte}
R.~Ebadi et~al., \emph{{Ultra-Heavy Dark Matter Search with Electron Microscopy
  of Geological Quartz}},  \href{http://arxiv.org/abs/2105.03998}{{\tt
  2105.03998}}.

\bibitem{Greiner:1990tz}
W.~Greiner, \emph{{Relativistic quantum mechanics: Wave equations}}.
\newblock 1990.

\bibitem{Migdal:1977bq}
A.~B. Migdal, \emph{{Qualitative Methods in Quantum Theory}}, vol.~48.
\newblock 1977.

\bibitem{Landau:1991wop}
L.~D. Landau and E.~Lifshits, \emph{{Quantum Mechanics}: {Non-Relativistic
  Theory}}, vol.~v.3 of \emph{Course of Theoretical Physics}.
\newblock Butterworth-Heinemann, Oxford, 1991.

\bibitem{Vergados:2004bm}
J.~D. Vergados and H.~Ejiri, \emph{{The role of ionization electrons in direct
  neutralino detection}},
  \href{http://dx.doi.org/10.1016/j.physletb.2004.11.085}{\emph{Phys. Lett. B}
  {\bf 606} (2005) 313--322}, [\href{http://arxiv.org/abs/hep-ph/0401151}{{\tt
  hep-ph/0401151}}].

\bibitem{Moustakidis:2005gx}
C.~Moustakidis, J.~Vergados and H.~Ejiri, \emph{{Direct dark matter detection
  by observing electrons produced in neutralino-nucleus collisions}},
  \href{http://dx.doi.org/10.1016/j.nuclphysb.2005.08.033}{\emph{Nucl. Phys. B}
  {\bf 727} (2005) 406--420}, [\href{http://arxiv.org/abs/hep-ph/0507123}{{\tt
  hep-ph/0507123}}].

\bibitem{Ejiri:2005aj}
H.~Ejiri, C.~Moustakidis and J.~Vergados, \emph{{Dark matter search by
  exclusive studies of X-rays following WIMPs nuclear interactions}},
  \href{http://dx.doi.org/10.1016/j.physletb.2006.03.037}{\emph{Phys. Lett. B}
  {\bf 639} (2006) 218--222}, [\href{http://arxiv.org/abs/hep-ph/0510042}{{\tt
  hep-ph/0510042}}].

\bibitem{Bernabei:2007jz}
R.~Bernabei et~al., \emph{{On electromagnetic contributions in WIMP quests}},
  \href{http://dx.doi.org/10.1142/S0217751X07037093}{\emph{Int. J. Mod. Phys.
  A} {\bf 22} (2007) 3155--3168}, [\href{http://arxiv.org/abs/0706.1421}{{\tt
  0706.1421}}].

\bibitem{Ibe:2017yqa}
M.~Ibe, W.~Nakano, Y.~Shoji and K.~Suzuki, \emph{{Migdal Effect in Dark Matter
  Direct Detection Experiments}},
  \href{http://dx.doi.org/10.1007/JHEP03(2018)194}{\emph{JHEP} {\bf 03} (2018)
  194}, [\href{http://arxiv.org/abs/1707.07258}{{\tt 1707.07258}}].

\bibitem{Dolan:2017xbu}
M.~J. Dolan, F.~Kahlhoefer and C.~McCabe, \emph{{Directly detecting sub-GeV
  dark matter with electrons from nuclear scattering}},
  \href{http://dx.doi.org/10.1103/PhysRevLett.121.101801}{\emph{Phys. Rev.
  Lett.} {\bf 121} (2018) 101801}, [\href{http://arxiv.org/abs/1711.09906}{{\tt
  1711.09906}}].

\bibitem{Essig:2019xkx}
R.~Essig, J.~Pradler, M.~Sholapurkar and T.-T. Yu, \emph{{Relation between the
  Migdal Effect and Dark Matter-Electron Scattering in Isolated Atoms and
  Semiconductors}},
  \href{http://dx.doi.org/10.1103/PhysRevLett.124.021801}{\emph{Phys. Rev.
  Lett.} {\bf 124} (2020) 021801}, [\href{http://arxiv.org/abs/1908.10881}{{\tt
  1908.10881}}].

\bibitem{Baxter:2019pnz}
D.~Baxter, Y.~Kahn and G.~Krnjaic, \emph{{Electron Ionization via Dark
  Matter-Electron Scattering and the Migdal Effect}},
  \href{http://dx.doi.org/10.1103/PhysRevD.101.076014}{\emph{Phys. Rev. D} {\bf
  101} (2020) 076014}, [\href{http://arxiv.org/abs/1908.00012}{{\tt
  1908.00012}}].

\bibitem{GrillidiCortona:2020owp}
G.~Grilli~di Cortona, A.~Messina and S.~Piacentini, \emph{{Migdal effect and
  photon Bremsstrahlung: improving the sensitivity to light dark matter of
  liquid argon experiments}},
  \href{http://dx.doi.org/10.1007/JHEP11(2020)034}{\emph{JHEP} {\bf 11} (2020)
  034}, [\href{http://arxiv.org/abs/2006.02453}{{\tt 2006.02453}}].

\bibitem{Liu:2020pat}
C.~P. Liu, C.-P. Wu, H.-C. Chi and J.-W. Chen, \emph{{Model-independent
  determination of the Migdal effect via photoabsorption}},
  \href{http://dx.doi.org/10.1103/PhysRevD.102.121303}{\emph{Phys. Rev. D} {\bf
  102} (2020) 121303}, [\href{http://arxiv.org/abs/2007.10965}{{\tt
  2007.10965}}].

\bibitem{Flambaum:2020xxo}
V.~V. Flambaum, L.~Su, L.~Wu and B.~Zhu, \emph{{Constraining sub-GeV dark
  matter from Migdal and Boosted effects}},
  \href{http://arxiv.org/abs/2012.09751}{{\tt 2012.09751}}.

\bibitem{Knapen:2020aky}
S.~Knapen, J.~Kozaczuk and T.~Lin, ``{The Migdal effect in semiconductors}.''
  11, 2020.

\bibitem{Bell:2021zkr}
N.~F. Bell, J.~B. Dent, B.~Dutta, S.~Ghosh, J.~Kumar and J.~L. Newstead,
  \emph{{Low-mass inelastic dark matter direct detection via the Migdal
  effect}},  \href{http://arxiv.org/abs/2103.05890}{{\tt 2103.05890}}.

\bibitem{Aprile:2017iyp}
{\scshape XENON} collaboration, E.~Aprile et~al., \emph{{First Dark Matter
  Search Results from the XENON1T Experiment}},
  \href{http://dx.doi.org/10.1103/PhysRevLett.119.181301}{\emph{Phys. Rev.
  Lett.} {\bf 119} (2017) 181301}, [\href{http://arxiv.org/abs/1705.06655}{{\tt
  1705.06655}}].

\bibitem{Iocco:2011jz}
F.~Iocco, M.~Pato, G.~Bertone and P.~Jetzer, \emph{{Dark Matter distribution in
  the Milky Way: microlensing and dynamical constraints}},
  \href{http://dx.doi.org/10.1088/1475-7516/2011/11/029}{\emph{JCAP} {\bf 11}
  (2011) 029}, [\href{http://arxiv.org/abs/1107.5810}{{\tt 1107.5810}}].

\bibitem{Acevedo:2020gro}
J.~F. Acevedo, J.~Bramante, A.~Goodman, J.~Kopp and T.~Opferkuch, \emph{{Dark
  Matter, Destroyer of Worlds: Neutrino, Thermal, and Existential Signatures
  from Black Holes in the Sun and Earth}},
  \href{http://dx.doi.org/10.1088/1475-7516/2021/04/026}{\emph{JCAP} {\bf 04}
  (2021) 026}, [\href{http://arxiv.org/abs/2012.09176}{{\tt 2012.09176}}].

\bibitem{Read:2014qva}
J.~I. Read, \emph{{The Local Dark Matter Density}},
  \href{http://dx.doi.org/10.1088/0954-3899/41/6/063101}{\emph{J. Phys. G} {\bf
  41} (2014) 063101}, [\href{http://arxiv.org/abs/1404.1938}{{\tt 1404.1938}}].

\bibitem{Pato:2015dua}
M.~Pato, F.~Iocco and G.~Bertone, \emph{{Dynamical constraints on the dark
  matter distribution in the Milky Way}},
  \href{http://dx.doi.org/10.1088/1475-7516/2015/12/001}{\emph{JCAP} {\bf 12}
  (2015) 001}, [\href{http://arxiv.org/abs/1504.06324}{{\tt 1504.06324}}].

\bibitem{beardsmore2001crustal}
G.~R. Beardsmore and J.~P. Cull, \emph{Crustal heat flow: a guide to
  measurement and modelling}.
\newblock Cambridge University Press, 2001.

\bibitem{Bramante:2018qbc}
J.~Bramante, B.~Broerman, R.~F. Lang and N.~Raj, \emph{{Saturated Overburden
  Scattering and the Multiscatter Frontier: Discovering Dark Matter at the
  Planck Mass and Beyond}},
  \href{http://dx.doi.org/10.1103/PhysRevD.98.083516}{\emph{Phys. Rev. D} {\bf
  98} (2018) 083516}, [\href{http://arxiv.org/abs/1803.08044}{{\tt
  1803.08044}}].

\bibitem{Bramante:2018tos}
J.~Bramante, B.~Broerman, J.~Kumar, R.~F. Lang, M.~Pospelov and N.~Raj,
  \emph{{Foraging for dark matter in large volume liquid scintillator neutrino
  detectors with multiscatter events}},
  \href{http://dx.doi.org/10.1103/PhysRevD.99.083010}{\emph{Phys. Rev. D} {\bf
  99} (2019) 083010}, [\href{http://arxiv.org/abs/1812.09325}{{\tt
  1812.09325}}].

\bibitem{Clark:2020mna}
M.~Clark, A.~Depoian, B.~Elshimy, A.~Kopec, R.~F. Lang and J.~Qin,
  \emph{{Direct Detection Limits on Heavy Dark Matter}},
  \href{http://dx.doi.org/10.1103/PhysRevD.102.123026}{\emph{Phys. Rev. D} {\bf
  102} (2020) 123026}, [\href{http://arxiv.org/abs/2009.07909}{{\tt
  2009.07909}}].

\bibitem{Bhoonah:2020fys}
A.~Bhoonah, J.~Bramante, B.~Courtman and N.~Song, \emph{{Etched plastic
  searches for dark matter}},
  \href{http://dx.doi.org/10.1103/PhysRevD.103.103001}{\emph{Phys. Rev. D} {\bf
  103} (2021) 103001}, [\href{http://arxiv.org/abs/2012.13406}{{\tt
  2012.13406}}].

\bibitem{Acevedo:2021tbl}
J.~F. Acevedo, J.~Bramante and A.~Goodman, \emph{{Old Rocks, New Limits:
  Excavated Ancient Mica Searches ForDark Matter}},
  \href{http://arxiv.org/abs/2105.06473}{{\tt 2105.06473}}.

\bibitem{Bernabei:1999ui}
R.~Bernabei et~al., \emph{{Extended limits on neutral strongly interacting
  massive particles and nuclearites from NaI(Tl) scintillators}},
  \href{http://dx.doi.org/10.1103/PhysRevLett.83.4918}{\emph{Phys. Rev. Lett.}
  {\bf 83} (1999) 4918--4921}.

\bibitem{Adhikari:2021fum}
P.~Adhikari et~al., \emph{{First direct detection constraints on Planck-scale
  mass dark matter with multiple-scatter signatures using the DEAP-3600
  detector}},  \href{http://arxiv.org/abs/2108.09405}{{\tt 2108.09405}}.

\bibitem{Hardy:2016kme}
E.~Hardy and R.~Lasenby, \emph{{Stellar cooling bounds on new light particles:
  plasma mixing effects}},
  \href{http://dx.doi.org/10.1007/JHEP02(2017)033}{\emph{JHEP} {\bf 02} (2017)
  033}, [\href{http://arxiv.org/abs/1611.05852}{{\tt 1611.05852}}].

\bibitem{Knapen:2017xzo}
S.~Knapen, T.~Lin and K.~M. Zurek, \emph{{Light Dark Matter: Models and
  Constraints}},
  \href{http://dx.doi.org/10.1103/PhysRevD.96.115021}{\emph{Phys. Rev. D} {\bf
  96} (2017) 115021}, [\href{http://arxiv.org/abs/1709.07882}{{\tt
  1709.07882}}].

\bibitem{Orito:1990ny}
S.~Orito et~al., \emph{{Search for supermassive relics with 2000-m**2 array of
  plastic track detector}},
  \href{http://dx.doi.org/10.1103/PhysRevLett.66.1951}{\emph{Phys. Rev. Lett.}
  {\bf 66} (1991) 1951--1954}.

\bibitem{Bramante:2019yss}
J.~Bramante, J.~Kumar and N.~Raj, \emph{{Dark matter astrometry at underground
  detectors with multiscatter events}},
  \href{http://dx.doi.org/10.1103/PhysRevD.100.123016}{\emph{Phys. Rev. D} {\bf
  100} (2019) 123016}, [\href{http://arxiv.org/abs/1910.05380}{{\tt
  1910.05380}}].

\bibitem{Scalzo:2014wxa}
R.~A. Scalzo, A.~J. Ruiter and S.~A. Sim, \emph{{The ejected mass distribution
  of type Ia supernovae: A significant rate of non-Chandrasekhar-mass
  progenitors}}, \href{http://dx.doi.org/10.1093/mnras/stu1808}{\emph{Mon. Not.
  Roy. Astron. Soc.} {\bf 445} (2014) 2535--2544},
  [\href{http://arxiv.org/abs/1408.6601}{{\tt 1408.6601}}].

\bibitem{NearbySupernovaFactory:2014mcg}
{\scshape Nearby Supernova Factory} collaboration, R.~Scalzo et~al.,
  \emph{{Type Ia supernova bolometric light curves and ejected mass estimates
  from the Nearby Supernova Factory}},
  \href{http://dx.doi.org/10.1093/mnras/stu350}{\emph{Mon. Not. Roy. Astron.
  Soc.} {\bf 440} (2014) 1498--1518},
  [\href{http://arxiv.org/abs/1402.6842}{{\tt 1402.6842}}].

\bibitem{1982ApJ...253..798N}
K.~{Nomoto}, \emph{{Accreting white dwarf models for type I supernovae. I -
  Presupernova evolution and triggering mechanisms}},
  \href{http://dx.doi.org/10.1086/159682}{\emph{The Astrophysical Journal
  Letters} {\bf 253} (Feb., 1982) 798--810}.

\bibitem{Han:2003uj}
Z.-W. Han and P.~Podsiadlowski, \emph{{The single degenerate channel for the
  progenitors of type Ia supernovae}},
  \href{http://dx.doi.org/10.1111/j.1365-2966.2004.07713.x}{\emph{Mon. Not.
  Roy. Astron. Soc.} {\bf 350} (2004) 1301},
  [\href{http://arxiv.org/abs/astro-ph/0309618}{{\tt astro-ph/0309618}}].

\bibitem{Wang:2018pac}
B.~Wang, \emph{{Mass-accreting white dwarfs and type Ia supernovae}},
  \href{http://dx.doi.org/10.1088/1674-4527/18/5/49}{\emph{Res. Astron.
  Astrophys.} {\bf 18} (2018) 049},
  [\href{http://arxiv.org/abs/1801.04031}{{\tt 1801.04031}}].

\bibitem{doi:10.1111/j.1365-2966.2011.19361.x}
A.~Kashi and N.~Soker, \emph{A circumbinary disc in the final stages of common
  envelope and the core-degenerate scenario for type ia supernovae},
  \href{http://dx.doi.org/10.1111/j.1365-2966.2011.19361.x}{\emph{Monthly
  Notices of the Royal Astronomical Society} {\bf 417} (2011) 1466--1479}.

\bibitem{doi:10.1093/mnras/stw1575}
D.-D. Liu, B.~Wang, P.~Podsiadlowski and Z.~Han, \emph{The violent white dwarf
  merger scenario for the progenitors of type ia supernovae},
  \href{http://dx.doi.org/10.1093/mnras/stw1575}{\emph{Monthly Notices of the
  Royal Astronomical Society} {\bf 461} (2016) 3653--3662}.

\bibitem{2041-8205-770-1-L8}
R.~Pakmor, M.~Kromer, S.~Taubenberger and V.~Springel, \emph{Helium-ignited
  violent mergers as a unified model for normal and rapidly declining type ia
  supernovae}, {\emph{The Astrophysical Journal Letters} {\bf 770} (2013) L8}.

\bibitem{2015Natur.521..332O}
R.~P. {Olling}, R.~{Mushotzky}, E.~J. {Shaya}, A.~{Rest}, P.~M. {Garnavich},
  B.~E. {Tucker} et~al., \emph{{No signature of ejecta interaction with a
  stellar companion in three type Ia supernovae}},
  \href{http://dx.doi.org/10.1038/nature14455}{\emph{Nature} {\bf 521} (May,
  2015) 332--335}.

\bibitem{Maoz:2013hna}
D.~Maoz, F.~Mannucci and G.~Nelemans, \emph{{Observational clues to the
  progenitors of Type-Ia supernovae}},
  \href{http://dx.doi.org/10.1146/annurev-astro-082812-141031}{\emph{Ann. Rev.
  Astron. Astrophys.} {\bf 52} (2014) 107--170},
  [\href{http://arxiv.org/abs/1312.0628}{{\tt 1312.0628}}].

\bibitem{Bramante:2015cua}
J.~Bramante, \emph{{Dark matter ignition of type Ia supernovae}},
  \href{http://dx.doi.org/10.1103/PhysRevLett.115.141301}{\emph{Phys. Rev.
  Lett.} {\bf 115} (2015) 141301}, [\href{http://arxiv.org/abs/1505.07464}{{\tt
  1505.07464}}].

\bibitem{Acevedo:2019gre}
J.~F. Acevedo and J.~Bramante, \emph{{Supernovae Sparked By Dark Matter in
  White Dwarfs}},
  \href{http://dx.doi.org/10.1103/PhysRevD.100.043020}{\emph{Phys. Rev. D} {\bf
  100} (2019) 043020}, [\href{http://arxiv.org/abs/1904.11993}{{\tt
  1904.11993}}].

\bibitem{Janish:2019nkk}
R.~Janish, V.~Narayan and P.~Riggins, \emph{{Type Ia supernovae from dark
  matter core collapse}},
  \href{http://dx.doi.org/10.1103/PhysRevD.100.035008}{\emph{Phys. Rev. D} {\bf
  100} (2019) 035008}, [\href{http://arxiv.org/abs/1905.00395}{{\tt
  1905.00395}}].

\bibitem{Graham:2018efk}
P.~W. Graham, R.~Janish, V.~Narayan, S.~Rajendran and P.~Riggins, \emph{{White
  Dwarfs as Dark Matter Detectors}},
  \href{http://dx.doi.org/10.1103/PhysRevD.98.115027}{\emph{Phys. Rev. D} {\bf
  98} (2018) 115027}, [\href{http://arxiv.org/abs/1805.07381}{{\tt
  1805.07381}}].

\bibitem{Fedderke:2019jur}
M.~A. Fedderke, P.~W. Graham and S.~Rajendran, \emph{{White dwarf bounds on
  charged massive particles}},
  \href{http://dx.doi.org/10.1103/PhysRevD.101.115021}{\emph{Phys. Rev. D} {\bf
  101} (2020) 115021}, [\href{http://arxiv.org/abs/1911.08883}{{\tt
  1911.08883}}].

\bibitem{Graham:2015apa}
P.~W. Graham, S.~Rajendran and J.~Varela, \emph{{Dark Matter Triggers of
  Supernovae}}, \href{http://dx.doi.org/10.1103/PhysRevD.92.063007}{\emph{Phys.
  Rev. D} {\bf 92} (2015) 063007}, [\href{http://arxiv.org/abs/1505.04444}{{\tt
  1505.04444}}].

\bibitem{Montero-Camacho:2019jte}
P.~Montero-Camacho, X.~Fang, G.~Vasquez, M.~Silva and C.~M. Hirata,
  \emph{{Revisiting constraints on asteroid-mass primordial black holes as dark
  matter candidates}},
  \href{http://dx.doi.org/10.1088/1475-7516/2019/08/031}{\emph{JCAP} {\bf 08}
  (2019) 031}, [\href{http://arxiv.org/abs/1906.05950}{{\tt 1906.05950}}].

\bibitem{Bertone:2007ae}
G.~Bertone and M.~Fairbairn, \emph{{Compact Stars as Dark Matter Probes}},
  \href{http://dx.doi.org/10.1103/PhysRevD.77.043515}{\emph{Phys. Rev. D} {\bf
  77} (2008) 043515}, [\href{http://arxiv.org/abs/0709.1485}{{\tt 0709.1485}}].

\bibitem{McCullough:2010ai}
M.~McCullough and M.~Fairbairn, \emph{{Capture of Inelastic Dark Matter in
  White Dwarves}},
  \href{http://dx.doi.org/10.1103/PhysRevD.81.083520}{\emph{Phys. Rev. D} {\bf
  81} (2010) 083520}, [\href{http://arxiv.org/abs/1001.2737}{{\tt 1001.2737}}].

\bibitem{Hooper:2010es}
D.~Hooper, D.~Spolyar, A.~Vallinotto and N.~Y. Gnedin, \emph{{Inelastic Dark
  Matter As An Efficient Fuel For Compact Stars}},
  \href{http://dx.doi.org/10.1103/PhysRevD.81.103531}{\emph{Phys. Rev. D} {\bf
  81} (2010) 103531}, [\href{http://arxiv.org/abs/1002.0005}{{\tt 1002.0005}}].

\bibitem{Amaro-Seoane:2015uny}
P.~Amaro-Seoane, J.~Casanellas, R.~Sch\"odel, E.~Davidson and J.~Cuadra,
  \emph{{Probing dark matter crests with white dwarfs and IMBHs}},
  \href{http://dx.doi.org/10.1093/mnras/stw433}{\emph{Mon. Not. Roy. Astron.
  Soc.} {\bf 459} (2016) 695--700},
  [\href{http://arxiv.org/abs/1512.00456}{{\tt 1512.00456}}].

\bibitem{Bramante:2017xlb}
J.~Bramante, A.~Delgado and A.~Martin, \emph{{Multiscatter stellar capture of
  dark matter}},
  \href{http://dx.doi.org/10.1103/PhysRevD.96.063002}{\emph{Phys. Rev. D} {\bf
  96} (2017) 063002}, [\href{http://arxiv.org/abs/1703.04043}{{\tt
  1703.04043}}].

\bibitem{Dasgupta:2019juq}
B.~Dasgupta, A.~Gupta and A.~Ray, \emph{{Dark matter capture in celestial
  objects: Improved treatment of multiple scattering and updated constraints
  from white dwarfs}},
  \href{http://dx.doi.org/10.1088/1475-7516/2019/08/018}{\emph{JCAP} {\bf 08}
  (2019) 018}, [\href{http://arxiv.org/abs/1906.04204}{{\tt 1906.04204}}].

\bibitem{Curtin:2020tkm}
D.~Curtin and J.~Setford, \emph{{Direct Detection of Atomic Dark Matter in
  White Dwarfs}}, \href{http://dx.doi.org/10.1007/JHEP03(2021)166}{\emph{JHEP}
  {\bf 03} (2021) 166}, [\href{http://arxiv.org/abs/2010.00601}{{\tt
  2010.00601}}].

\bibitem{Horowitz:2020axx}
C.~J. Horowitz, \emph{{Nuclear and dark matter heating in massive white dwarf
  stars}}, \href{http://dx.doi.org/10.1103/PhysRevD.102.083031}{\emph{Phys.
  Rev. D} {\bf 102} (2020) 083031},
  [\href{http://arxiv.org/abs/2008.03291}{{\tt 2008.03291}}].

\bibitem{Chan:2020ijt}
H.-S. Chan, M.-C. Chu, S.-C. Leung and L.-M. Lin, \emph{{Delayed Detonation
  Thermonuclear Supernovae With An Extended Dark Matter Component}},
  \href{http://arxiv.org/abs/2012.06857}{{\tt 2012.06857}}.

\bibitem{Dasgupta:2020mqg}
B.~Dasgupta, R.~Laha and A.~Ray, \emph{{Low Mass Black Holes from Dark Core
  Collapse}},
  \href{http://dx.doi.org/10.1103/PhysRevLett.126.141105}{\emph{Phys. Rev.
  Lett.} {\bf 126} (2021) 141105}, [\href{http://arxiv.org/abs/2009.01825}{{\tt
  2009.01825}}].

\bibitem{Bell:2021fye}
N.~F. Bell, G.~Busoni, M.~E. Ramirez-Quezada, S.~Robles and M.~Virgato,
  \emph{{Improved Treatment of Dark Matter Capture in White Dwarfs}},
  \href{http://arxiv.org/abs/2104.14367}{{\tt 2104.14367}}.

\bibitem{1992Timmes}
F.~X. {Timmes} and S.~E. {Woosley}, \emph{{The conductive propagation of
  nuclear flames. I - Degenerate C $+$ O and O $+$ NE $+$ MG white dwarfs}},
  \href{http://dx.doi.org/10.1086/171746}{\emph{Astrophys. J.} {\bf 396}
  (Sept., 1992) 649--667}.

\bibitem{Potekhin:1999yv}
A.~Potekhin, D.~Baiko, P.~Haensel and D.~Yakovlev, \emph{{Transport properties
  of degenerate electrons in neutron star envelopes and white dwarf cores}},
  {\emph{Astron. Astrophys.} {\bf 346} (1999) 345},
  [\href{http://arxiv.org/abs/astro-ph/9903127}{{\tt astro-ph/9903127}}].

\bibitem{1983bhwd.book.....S}
S.~L. {Shapiro} and S.~A. {Teukolsky}, \emph{{Black holes, white dwarfs, and
  neutron stars: The physics of compact objects}}.
\newblock 1983.

\bibitem{Kippenhahn:1994wva}
R.~Kippenhahn, A.~Weigert and A.~Weiss, \emph{{Stellar structure and
  evolution}}, vol.~9783642303043.
\newblock Springer, 8, 2012,
  \href{http://dx.doi.org/10.1007/978-3-642-30304-3}{10.1007/978-3-642-30304-3}.

\bibitem{1987ApJ...313..531S}
P.~J. {Schinder}, D.~N. {Schramm}, P.~J. {Wiita}, S.~H. {Margolis} and D.~L.
  {Tubbs}, \emph{{Neutrino Emission by the Pair, Plasma, and Photo Processes in
  the Weinberg-Salam Model}},
  \href{http://dx.doi.org/10.1086/164993}{\emph{Astrophysical Journal} {\bf
  313} (Feb., 1987) 531}.

\bibitem{1989ApJ...339..354I}
N.~{Itoh}, T.~{Adachi}, M.~{Nakagawa}, Y.~{Kohyama} and H.~{Munakata},
  \emph{{Neutrino Energy Loss in Stellar Interiors. III. Pair, Photo-, Plasma,
  and Bremsstrahlung Processes}},
  \href{http://dx.doi.org/10.1086/167301}{\emph{Astrophysical Journal} {\bf
  339} (Apr., 1989) 354}.

\bibitem{1996ApJS..102..411I}
N.~{Itoh}, H.~{Hayashi}, A.~{Nishikawa} and Y.~{Kohyama}, \emph{{Neutrino
  Energy Loss in Stellar Interiors. VII. Pair, Photo-, Plasma, Bremsstrahlung,
  and Recombination Neutrino Processes}},
  \href{http://dx.doi.org/10.1086/192264}{\emph{Astrophysical Journal
  Supplement} {\bf 102} (Feb., 1996) 411}.

\bibitem{Gasques:2005ar}
L.~Gasques, A.~Afanasjev, E.~Aguilera, M.~Beard, L.~Chamon, P.~Ring et~al.,
  \emph{{Nuclear fusion in dense matter: Reaction rate and carbon burning}},
  \href{http://dx.doi.org/10.1103/PhysRevC.72.025806}{\emph{Phys. Rev. C} {\bf
  72} (2005) 025806}, [\href{http://arxiv.org/abs/astro-ph/0506386}{{\tt
  astro-ph/0506386}}].

\bibitem{Yakovlev:2006fi}
D.~g. Yakovlev, L.~R. Gasques, M.~Beard, M.~Wiescher and A.~V. Afanasjev,
  \emph{{Fusion reactions in multicomponent dense matter}},
  \href{http://dx.doi.org/10.1103/PhysRevC.74.035803}{\emph{Phys. Rev. C} {\bf
  74} (2006) 035803}, [\href{http://arxiv.org/abs/astro-ph/0608488}{{\tt
  astro-ph/0608488}}].

\bibitem{Caughlan:1987qf}
G.~R. Caughlan and W.~A. Fowler, \emph{{Thermonuclear reaction rates. 5.}},
  \href{http://dx.doi.org/10.1016/0092-640X(88)90009-5}{\emph{Atom. Data Nucl.
  Data Tabl.} {\bf 40} (1988) 283--334}.

\bibitem{2017A&A...602A..16T}
S.~{Toonen}, M.~{Hollands}, B.~T. {G{\"a}nsicke} and T.~{Boekholt}, \emph{{The
  binarity of the local white dwarf population}},
  \href{http://dx.doi.org/10.1051/0004-6361/201629978}{\emph{Astronomy \&
  Astrophysics} {\bf 602} (June, 2017) A16},
  [\href{http://arxiv.org/abs/1703.06893}{{\tt 1703.06893}}].

\bibitem{Maoz:2018epf}
D.~Maoz, N.~Hallakoun and C.~Badenes, \emph{{The separation distribution and
  merger rate of double white dwarfs: improved constraints}},
  \href{http://dx.doi.org/10.1093/mnras/sty339}{\emph{Mon. Not. Roy. Astron.
  Soc.} {\bf 476} (2018) 2584--2590},
  [\href{http://arxiv.org/abs/1801.04275}{{\tt 1801.04275}}].

\bibitem{2020A&A...636A..31T}
K.~D. {Temmink}, S.~{Toonen}, E.~{Zapartas}, S.~{Justham} and B.~T.
  {G{\"a}nsicke}, \emph{{Looks can be deceiving. Underestimating the age of
  single white dwarfs due to binary mergers}},
  \href{http://dx.doi.org/10.1051/0004-6361/201936889}{\emph{Astronomy \&
  Astrophysics} {\bf 636} (Apr., 2020) A31},
  [\href{http://arxiv.org/abs/1910.05335}{{\tt 1910.05335}}].

\bibitem{Yoon:2007pw}
S.-C. Yoon, P.~Podsiadlowski and S.~Rosswog, \emph{{Remnant evolution after a
  carbon-oxygen white dwarf merger}},
  \href{http://dx.doi.org/10.1111/j.1365-2966.2007.12161.x}{\emph{Mon. Not.
  Roy. Astron. Soc.} {\bf 380} (2007) 933},
  [\href{http://arxiv.org/abs/0704.0297}{{\tt 0704.0297}}].

\bibitem{LorenAguilar:2009cv}
P.~Loren-Aguilar, J.~Isern and E.~Garcia-Berro, \emph{{High-resolution Smoothed
  Particle Hydrodynamics simulations of the merger of binary white dwarfs}},
  \href{http://dx.doi.org/10.1063/1.3141311}{\emph{AIP Conf. Proc.} {\bf 1122}
  (2009) 320--323}, [\href{http://arxiv.org/abs/0903.4599}{{\tt 0903.4599}}].

\bibitem{2019NatAs...3..408D}
L.~{Decin}, W.~{Homan}, T.~{Danilovich}, A.~{de Koter}, D.~{Engels},
  L.~B.~F.~M. {Waters} et~al., \emph{{Reduction of the maximum mass-loss rate
  of OH/IR stars due to unnoticed binary interaction}},
  \href{http://dx.doi.org/10.1038/s41550-019-0703-5}{\emph{Nature Astronomy}
  {\bf 3} (Feb., 2019) 408--415}, [\href{http://arxiv.org/abs/1902.09259}{{\tt
  1902.09259}}].

\bibitem{2021A&A...646A..30A}
L.~G. {Althaus}, P.~{Gil-Pons}, A.~H. {C{\'o}rsico}, M.~{Miller Bertolami},
  F.~{De Ger{\'o}nimo}, M.~E. {Camisassa} et~al., \emph{{The formation of
  ultra-massive carbon-oxygen core white dwarfs and their evolutionary and
  pulsational properties}},
  \href{http://dx.doi.org/10.1051/0004-6361/202038930}{\emph{Astronomy \&
  Astrophysics} {\bf 646} (Feb., 2021) A30},
  [\href{http://arxiv.org/abs/2011.10439}{{\tt 2011.10439}}].

\bibitem{1996ApJ...472..783D}
I.~{Dominguez}, O.~{Straniero}, A.~{Tornambe} and J.~{Isern}, \emph{{On the
  Formation of Massive C-O White Dwarfs: The Lifting Effect of Rotation}},
  \href{http://dx.doi.org/10.1086/178106}{\emph{Astrophysical Journal} {\bf
  472} (Dec., 1996) 783}.

\bibitem{2017ASPC}
P.~{Dufour}, S.~{Blouin}, S.~{Coutu}, M.~{Fortin-Archambault}, C.~{Thibeault},
  P.~{Bergeron} et~al., \emph{{The Montreal White Dwarf Database: A Tool for
  the Community}},  in \emph{20th European White Dwarf Workshop} (P.-E.
  {Tremblay}, B.~{Gaensicke} and T.~{Marsh}, eds.), vol.~509 of
  \emph{Astronomical Society of the Pacific Conference Series}, p.~3, Mar.,
  2017.
\newblock \href{http://arxiv.org/abs/1610.00986}{{\tt 1610.00986}}.

\bibitem{Kawasaki:1991eu}
M.~Kawasaki, H.~Murayama and T.~Yanagida, \emph{{Can the strongly interacting
  dark matter be a heating source of Jupiter?}},
  \href{http://dx.doi.org/10.1143/PTP.87.685}{\emph{Prog. Theor. Phys.} {\bf
  87} (1992) 685--692}.

\bibitem{Mitra:2004fh}
S.~Mitra, \emph{{Uranus' anomalously low excess heat constrains strongly
  interacting dark matter}},
  \href{http://dx.doi.org/10.1103/PhysRevD.70.103517}{\emph{Phys. Rev. D} {\bf
  70} (2004) 103517}, [\href{http://arxiv.org/abs/astro-ph/0408341}{{\tt
  astro-ph/0408341}}].

\bibitem{Mack:2007xj}
G.~D. Mack, J.~F. Beacom and G.~Bertone, \emph{{Towards Closing the Window on
  Strongly Interacting Dark Matter: Far-Reaching Constraints from Earth's Heat
  Flow}}, \href{http://dx.doi.org/10.1103/PhysRevD.76.043523}{\emph{Phys. Rev.
  D} {\bf 76} (2007) 043523}, [\href{http://arxiv.org/abs/0705.4298}{{\tt
  0705.4298}}].

\bibitem{Adler:2008ky}
S.~L. Adler, \emph{{Planet-bound dark matter and the internal heat of Uranus,
  Neptune, and hot-Jupiter exoplanets}},
  \href{http://dx.doi.org/10.1016/j.physletb.2008.12.023}{\emph{Phys. Lett. B}
  {\bf 671} (2009) 203--206}, [\href{http://arxiv.org/abs/0808.2823}{{\tt
  0808.2823}}].

\bibitem{Chauhan:2016joa}
B.~Chauhan and S.~Mohanty, \emph{{Constraints on leptophilic light dark matter
  from internal heat flux of Earth}},
  \href{http://dx.doi.org/10.1103/PhysRevD.94.035024}{\emph{Phys. Rev. D} {\bf
  94} (2016) 035024}, [\href{http://arxiv.org/abs/1603.06350}{{\tt
  1603.06350}}].

\bibitem{Bramante:2019fhi}
J.~Bramante, A.~Buchanan, A.~Goodman and E.~Lodhi, \emph{{Terrestrial and
  Martian Heat Flow Limits on Dark Matter}},
  \href{http://dx.doi.org/10.1103/PhysRevD.101.043001}{\emph{Phys. Rev. D} {\bf
  101} (2020) 043001}, [\href{http://arxiv.org/abs/1909.11683}{{\tt
  1909.11683}}].

\bibitem{Garani:2019rcb}
R.~Garani and P.~Tinyakov, \emph{{Constraints on Dark Matter from the Moon}},
  \href{http://dx.doi.org/10.1016/j.physletb.2020.135403}{\emph{Phys. Lett. B}
  {\bf 804} (2020) 135403}, [\href{http://arxiv.org/abs/1912.00443}{{\tt
  1912.00443}}].

\bibitem{Chan:2020vsr}
M.~H. Chan and C.~M. Lee, \emph{{Constraining the spin-independent elastic
  scattering cross section of dark matter using the Moon as a detection target
  and the background neutrino data}},
  \href{http://dx.doi.org/10.1103/PhysRevD.102.023024}{\emph{Phys. Rev. D} {\bf
  102} (2020) 023024}, [\href{http://arxiv.org/abs/2007.01589}{{\tt
  2007.01589}}].

\bibitem{Leane:2020wob}
R.~K. Leane and J.~Smirnov, \emph{{Exoplanets as Sub-GeV Dark Matter
  Detectors}},
  \href{http://dx.doi.org/10.1103/PhysRevLett.126.161101}{\emph{Phys. Rev.
  Lett.} {\bf 126} (2021) 161101}, [\href{http://arxiv.org/abs/2010.00015}{{\tt
  2010.00015}}].

\bibitem{Leane:2021tjj}
R.~K. Leane and T.~Linden, \emph{{First Analysis of Jupiter in Gamma Rays and a
  New Search for Dark Matter}},  \href{http://arxiv.org/abs/2104.02068}{{\tt
  2104.02068}}.

\bibitem{williams1974}
D.~L. Williams, R.~P. Von~Herzen, J.~G. Sclater and R.~N. Anderson, \emph{{The
  Galapagos Spreading Centre: Lithospheric Cooling and Hydrothermal
  Circulation*}},
  \href{http://dx.doi.org/10.1111/j.1365-246X.1974.tb05431.x}{\emph{Geophysical
  Journal International} {\bf 38} (09, 1974) 587--608}.

\bibitem{lister1990}
C.~R.~B. Lister, J.~G. Sclater, E.~E. Davis, H.~Villinger and S.~Nagihara,
  \emph{{Heat flow maintained in ocean basins of great age: investigations in
  the north-equatorial West Pacific}},
  \href{http://dx.doi.org/10.1111/j.1365-246X.1990.tb04586.x}{\emph{Geophysical
  Journal International} {\bf 102} (09, 1990) 603--630}.

\bibitem{Buch:2018qdr}
J.~Buch, S.~C.~J. Leung and J.~Fan, \emph{{Using Gaia DR2 to Constrain Local
  Dark Matter Density and Thin Dark Disk}},
  \href{http://dx.doi.org/10.1088/1475-7516/2019/04/026}{\emph{JCAP} {\bf 04}
  (2019) 026}, [\href{http://arxiv.org/abs/1808.05603}{{\tt 1808.05603}}].

\bibitem{Gresham:2018rqo}
M.~I. Gresham and K.~M. Zurek, \emph{{Asymmetric Dark Stars and Neutron Star
  Stability}}, \href{http://dx.doi.org/10.1103/PhysRevD.99.083008}{\emph{Phys.
  Rev. D} {\bf 99} (2019) 083008}, [\href{http://arxiv.org/abs/1809.08254}{{\tt
  1809.08254}}].

\bibitem{Coskuner:2018are}
A.~Coskuner, D.~M. Grabowska, S.~Knapen and K.~M. Zurek, \emph{{Direct
  Detection of Bound States of Asymmetric Dark Matter}},
  \href{http://dx.doi.org/10.1103/PhysRevD.100.035025}{\emph{Phys. Rev. D} {\bf
  100} (2019) 035025}, [\href{http://arxiv.org/abs/1812.07573}{{\tt
  1812.07573}}].

\bibitem{Helm:1956zz}
R.~H. Helm, \emph{{Inelastic and Elastic Scattering of 187-Mev Electrons from
  Selected Even-Even Nuclei}},
  \href{http://dx.doi.org/10.1103/PhysRev.104.1466}{\emph{Phys. Rev.} {\bf 104}
  (1956) 1466--1475}.

\bibitem{Lewin:1995rx}
J.~Lewin and P.~Smith, \emph{{Review of mathematics, numerical factors, and
  corrections for dark matter experiments based on elastic nuclear recoil}},
  \href{http://dx.doi.org/10.1016/S0927-6505(96)00047-3}{\emph{Astropart.
  Phys.} {\bf 6} (1996) 87--112}.

\bibitem{Grabowska:2018lnd}
D.~M. Grabowska, T.~Melia and S.~Rajendran, \emph{{Detecting Dark Blobs}},
  \href{http://dx.doi.org/10.1103/PhysRevD.98.115020}{\emph{Phys. Rev. D} {\bf
  98} (2018) 115020}, [\href{http://arxiv.org/abs/1807.03788}{{\tt
  1807.03788}}].

\bibitem{Joglekar:2019vzy}
A.~Joglekar, N.~Raj, P.~Tanedo and H.-B. Yu, \emph{{Relativistic capture of
  dark matter by electrons in neutron stars}},
  \href{http://dx.doi.org/10.1016/j.physletb.2020.135767}{\emph{Phys. Lett.}
  {\bf B} (2020) 135767}, [\href{http://arxiv.org/abs/1911.13293}{{\tt
  1911.13293}}].

\bibitem{Joglekar:2020liw}
A.~Joglekar, N.~Raj, P.~Tanedo and H.-B. Yu, \emph{{Kinetic Heating from
  Contact Interactions with Relativistic Targets: Electrons Capture Dark Matter
  in Neutron Stars}},  \href{http://arxiv.org/abs/2004.09539}{{\tt
  2004.09539}}.

\bibitem{landau2013classical}
L.~Landau, \emph{The Classical Theory of Fields}.
\newblock Course of theoretical physics. Elsevier Science, 2013.

\bibitem{Cannoni:2016hro}
M.~Cannoni, \emph{{Lorentz invariant relative velocity and relativistic binary
  collisions}}, \href{http://dx.doi.org/10.1142/S0217751X17300022}{\emph{Int.
  J. Mod. Phys. A} {\bf 32} (2017) 1730002},
  [\href{http://arxiv.org/abs/1605.00569}{{\tt 1605.00569}}].

\end{thebibliography}\endgroup

\end{document}